\documentclass[fleqn,usenatbib]{mnras}

\usepackage{amssymb}	
\usepackage{newtxtext,newtxmath}

\usepackage[T1]{fontenc}
\usepackage[utf8]{inputenc}

\DeclareRobustCommand{\VAN}[3]{#2}
\let\VANthebibliography\thebibliography
\def\thebibliography{\DeclareRobustCommand{\VAN}[3]{##3}\VANthebibliography}

\usepackage{graphicx}	
\usepackage{enumitem}
\usepackage{booktabs,tabularx,graphicx,amsmath,makecell,hhline,enumitem,gensymb,multirow,ulem,adjustbox}
\usepackage{hyperref}
\usepackage{bbding,pifont}
\usepackage{hyperref} 
\usepackage[dvipsnames]{xcolor}
\usepackage{dcolumn} 
\newcommand{\mycomment}[1]{}

\usepackage{tikz}
\usetikzlibrary{shapes.geometric, shapes.symbols}

\title[$\omega$Cen and the GSE walk into a bar]{
Stirred, not shaken: Dislodging $\omega$ Centauri from the Sausage galaxy through bar resonances }

\author[A. Mannu et al.]{Axel Mannu$^{1}$, Chervin F. P. Laporte$^{1,2,3,4}$\thanks{E-mail: chervin.laporte@obpsm.fr}, Matthew D. A. Orkney,$^{2,3}$,  Yassin Rany Khalil$^{5}$ and Benoit Famaey$^{6}$ \vspace{0.1cm}\\
$^{1}$LIRA, Observatoire de Paris, Universit\'e PSL, Sorbonne Universit\'e, Universit\'e Paris Cit\'e, CY Cergy Paris Universit\'e, CNRS, 92190 Meudon, France\\
$^{2}$Institut de Ci\`{e}ncies del Cosmos (ICCUB), Universitat de Barcelona, Mart\'{i} i Franqu\`{e}s 1, E-08028 Barcelona, Spain\\
$^{3}$Institut d'Estudis Espacials de Catalunya (IEEC), E-08034 Barcelona, Spain\\
$^{4}$Kavli IPMU (WPI), UTIAS, The University of Tokyo, Kashiwa, Chiba 277-8583, Japan\\
$^{5}$Dipartimento di Fisica e Astronomia, Universit\`a degli Studi di Firenze, Via G. Sansone 1, I-50019 Sesto Fiorentino, Italy\\
$^{6}$Université de Strasbourg, CNRS, Observatoire astronomique de Strasbourg, UMR 7550, F-67000 Strasbourg, France\\
}

\date{Accepted XXX. Received YYY; in original form ZZZ}

\pubyear{2025}

\begin{document}
\label{firstpage}
\pagerange{\pageref{firstpage}--\pageref{lastpage}}
\maketitle

\begin{abstract}

We test the idea proposed by Laporte \& Orkney that $\omega$ Centauri ($\omega$Cen) is the central nuclear star cluster of the Gaia-Sausage-Enceladus  galaxy (GSE) which got dislodged by the bar. We run a series of simulations in time-evolving bar potentials with present-day pattern speeds within Portail et al's $\Omega_{b}=39\pm3.5\,\rm{km\,s^{-1}\,kpc^{-1}}$ and rate of slow-down $\dot{\Omega}_{\rm b}= 4.5 \pm 1.4 \,\rm{km \,s^{-1}\,kpc^{-1}\, Gyr^{-1}}$ to study how bar resonances modulate the angular momentum and energies of both GSE-debris and $\omega$Cen's location in $(E,L_{z})$ space. We find that the retrograde $3:2$ resonance is capable of inflicting changes of $\Delta E\approx0.4\times10^{5}\,\rm{km^{2}\, s^{-2}}$, $\Delta L_{z}\approx0.3-0.5\times10^{3}\, \rm{km \, s^{-1}\, kpc^{-1}}$ within timescales of $\Delta t\approx 0.5-1\,\rm{Gyr}$, well within the bar's lifespan $t_{\rm{bar}}\approx10-8\,\rm{Gyr}$. These results are independent of the choice of potential or modelling. We uncover a sample of N-rich stars with metallicities $-2.<[\rm{Fe/H}]<-1.0$ and aluminium enrichments consistent with $\omega$Cen. In ($E,L_{z}$) space, the map reveals a corridor of trailing N-rich stars clustered about the region connecting $\omega$Cen to the GSE's debris centroid, matching $\omega$Cen's past predicted track. This provides the best evidence yet that $\omega$Cen is the nuclear star cluster of the GSE. Using M54 and $\omega$Cen's  chemistry we show that neither first nor second generations follow general galaxy-wide scale chemical enrichment trends. This is empirical evidence that GC chemical trends are a cluster-scale physics problem, not a galaxy-wide one.

\end{abstract}




\section{Introduction} \label{sec:introduction}

Within the current cosmological paradigm \citep{Planck2014}, the Milky Way (MW) is thought to have formed through gravitational collapse of the dark matter density field, giving rise to grand structures known as dark matter haloes \citep[e.g.][]{efstathiou1985}. These haloes grow hierarchically through consecutive mergers, producing the deep gravitational wells that facilitate the condensation of baryons and ultimately give rise to galaxy formation \citep{white1978,white1991,springel2005}. 

For decades, there have been clues that the MW underwent a significant merger early in its formation history. One of the first indications came from the discovery that a large fraction of old, metal-poor halo stars occupy highly eccentric orbits \citep{chiba2000}. Although this was initially interpreted as the signature of a rapid, monolithic collapse of proto-Galactic gas as in \citet{eggen1962}, subsequent chemodynamical \citep{brook2003} and cosmological \citep{meza2005} simulations demonstrated that the accretion of dwarf galaxies provides a more natural explanation. Additional support came from the identification of a break in the stellar halo density profile at $\sim25\,$kpc \citep{watkins2009, sesar2011, deason2011}, interpreted as the accumulation of tidal debris near orbital apocentre following a massive and early accretion event \citep{deason2013, pillepich2014, amorisco2017}. In parallel, high-resolution spectroscopic studies of nearby stars \citep[e.g.][]{nissen2010} showed that halo stars definitely fall into two distinct populations, clearly separated in $[\rm{\alpha/Fe}]$.

The advent of \textit{Gaia} DR1 confirmed the galactic accretion scenario through the discovery of a massive, radially anisotropic stellar population that contributes up to half of the local stellar halo \citep{belokurov2018}. This population was associated with a single massive progenitor, now known as the \textit{Gaia} Sausage Enceladus (GSE). Independent support for the GSE has since emerged from the identification of distinct sequences in the stellar halo colour--magnitude diagram \citep{gaia2018, haywood2018, helmi2018} and a bimodal stellar age distribution \citep[e.g.][]{gallart2019}.

The discovery of this major Galactic contributor to the stellar halo (which accounts for less than 2 percent of the total stellar mass of the Galaxy) has launched a search for the debris of the remaining ancient accretion events, leading to an ever-expanding inventory of candidate mergers identified in various chemo-dynamical spaces \citep[e.g.][]{matsuno2019, koppelman2019, myeong19, naidu2020, malhan2022, lovdal2022, myeong2022, dodd2023}. The integrals of motion, particularly the orbital energy and $z$-component of the angular momentum, have been widely used due to the expectation that they are approximately conserved \citep[see][]{helmi2000, binney2008}, and their capacity to retain information about the masses and orbital properties of the mergers themselves \citep[e.g.][]{amorisco2017}. However, there has been growing awareness as to the limitations of this approach \citep[i.e.][]{amarante2022, khoperskov2023, mori2024, thomas2025}, raising questions as to whether some identified substructures belong to the same progenitor galaxy.


On the other hand, the globular cluster (GC) $\omega$ Centauri, hereafter $\omega$Cen, is the most massive cluster in the MW with $M_{\star}\approx4.5\times10^6\,{\rm M}{\odot}$ \citep{dsouza2013}. $\omega$Cen is an exceptional GC in many regards, featuring a large metallicity spread indicative of prolonged star formation in a dense environment \citep{freeman1975, smith2000, johnson2010}, and at least three distinct stellar populations that likely owe to independent formation mechanisms \citep[commonly referred to as the P1, Intermediate, and P2 populations,][]{woolley1966, lee1999, bedin2004, gratton2011, mason2026}. Due to these oddities, it has long been regarded as the remnant nuclear star cluster (NSC) of an accreted dwarf galaxy \citep[e.g.][]{zinnecker1988, lee1999, majewski2000, hilker2000, bekki2003}.

Even before the GSE was formally identified, \citet{meza2005} argued that $\omega$Cen and a population of metal-poor halo stars on highly eccentric orbits shared a common origin based on their specific angular momenta and chemistry. Now, several studies have argued that $\omega$Cen is among the GCs most strongly associated with the GSE \citep[e.g.][]{massari2019, pfeffer2021} albeit with a number of challenges \citep[see][]{pfeffer2021} . Consistent with this interpretation, \citet{limberg2022} estimated the stellar mass of $\omega$Cen's progenitor to be $M_{\star}\approx1.3\times10^9\,{\rm M}_\odot$, within the range commonly inferred for the GSE progenitor galaxy \citep[e.g.][]{helmi2018, alvar2018, mackereth2019, vincenzo2019, feuillet2020, naidu2020, kruijssen2020, han2022, lane2023}. 
\citet{souza2026} push the scenario further with chemodynamical arguments linking $\omega$Cen, \textit{Thamnos}, GSE, and \textit{Sequoia} to the shredded debris of a single progenitor galaxy which they call the Omega dwarf, though this interpretation remains highly tentative. Recently,
\cite{laporte2026} combined both dynamical and chemical arguments to show that the GCs associated with the GSE follow the same age-metallicity relation as that of $\omega$Cen's stellar populations as measured by \cite{clontz24}.

Conversely, there are compelling reasons against linking $\omega$Cen to the GSE. Firstly, its complex chemical enrichment differs from that of typical GSE stars \citep[e.g.][]{anguiano2025, mason2026}, though this may be justified due to the complex interplay of self-enrichment and external gas accretion within $\omega$Cen \citep[e.g.][]{bekki2003, bekki2006}. Secondly, its retrograde orbit does not reflect the highly radialised debris of the GSE, leading \citet{myeong19} to associate it instead with the retrograde \textit{Sequoia} halo substructure. Alternatively, \citet{pagnini2025,pagnini2026} argue that $\omega$Cen may originate from a yet undiscovered progenitor which they name \textit{Nephele}. However, invoking further massive merger events in the MW's early assembly comes up against a theoretical limit based on hierarchical assembly in the $\Lambda$CDM paradigm: the majority of early, accreted stars should owe to just 1--3 accretion events \citep[see arguments in][]{robertson05, fattahi20}. 

An important complication is that not all orbits have well-conserved integrals of motion. Cumulative random interactions with Galactic structures like the bar lead to a diffusion in both energy and angular momentum \citep{dillamore2025}, and interactions with bar resonances can cause systematic movement or clustering in the ($E,L_{\rm z}$) parameter space \citep{dillamore2023, dillamore2024b, tomlinson26, deleo2026, gherghinescu2026}. Furthermore, interactions with satellite substructures can scatter the orbits of both stars and GCs \citep{jean-baptiste17, garrow2020, amarante2022, pagnini2023, orkney26}, meaning that the present-day integrals of motion are not necessarily a reliable metric to judge past associations.

\citet{laporte2026} were the first to link these effects to $\omega$Cen and argue for a common origin. In particular, they proposed that interactions with bar resonances may have transported the cluster towards lower orbital energies and more retrograde angular momenta, thereby reconciling its apparent kinematic mismatch with the GSE stellar debris. This is plausible given that the Galactic bar is believed to have formed relatively soon after the GSE merger (e.g. $\sim8\,\mathrm{Gyr}$ ago, \citealp{bovy2019}; $\sim9\,\mathrm{Gyr}$ ago, \citealp{grady2020}; $8\text{--}10\,\mathrm{Gyr}$ ago, \citealp{haywood2024}), and may even owe its formation to the merger event \citep{merrow2024}.

If this interpretation is correct, it would resolve several long-standing puzzles surrounding $\omega$Cen. Firstly, it would remove the need to invoke an additional massive merger in the early assembly of the MW. Secondly, it would provide a mechanism for producing the strongly retrograde orbit. Furthermore, if the bar formed around the same time as the accretion of the GSE and $\omega$Cen, there would have been ample time for resonant interactions to alter the cluster's orbit. Independent support for this scenario may come from metallicity ridges that develop in the ($E,L_{\rm z}$) plane, which arise due to the resonance-induced migration of halo stars with a pre-existing metallicity gradient \citep{tomlinson26}. Interestingly, \citet{laporte2026} found evidence for such a feature in the GSE debris: their figure~5 shows a `corridor' of halo stars in the ($E,L_{\rm z}$) plane that bridges the gap between the core of the GSE footprint and $\omega$Cen's orbit, with a relative metal-enhancement of $\Delta\text{[Fe/H]}\approx0.25$. This was corroborated by \citet{dillamore2026} although further confirmation would need to await high-resolution spectroscopy.

\citet{dillamore2026} explored the possibility of a migration of $\omega$Cen caused by the $1:1$ retrograde resonance of the bar, which necessarily implies a very slow pattern speed, with a present-day value of $ \Omega_{\rm b} \lesssim 26\,\rm{km \,s^{-1}\, kpc^{-1}}$, consistent with the estimate in \citet{horta25}. This value is however at odds with most other studies, favouring a much higher pattern speed, such as that of \citet{portail17}, who infer $\Omega_{\rm bar}=39\pm 3.5 \,\rm{km \,s^{-1}\, kpc^{-1}}$, corroborated by, e.g., the recent study of \citet{khalil25} who fit a pattern speed of $37\, \rm{ km \, s^{-1} \,kpc^{-1}}$ to \textit{Gaia} DR3 data using a model that jointly includes the spiral arms. However, it is highly likely that the bar had a higher pattern speed at birth \citep[see e.g.][]{chiba21}, leaving room for numerous bar resonances to have impacted both the GSE debris and $\omega$Cen in the past.

In this work, we ask whether orbital resonances with a stellar bar displaying a realistic present-day pattern speed can plausibly and efficiently dislodge $\omega$Cen from an orbit whose energy and angular momentum were initially aligned with those of GSE halo stars. More specifically, we investigate which resonances besides the $1:1$ retrograde resonance could have been important for this mechanism.  In section~\ref{sec:sims} we describe our numerical simulations used to model the effect of the bar on tracers in the stellar halo (GSE debris, $\omega$Cen). In section~\ref{sec:stellar-halo} we search the stellar halo through high-resolution spectroscopic data for observational clues on the relation between between host galaxy and NSC and a new smoking gun signature on the link between $\omega$Cen and GSE. We discuss our findings in the context of the formation of the MW in section~\ref{sec:discussion} and conclude in section~\ref{sec:conclusion}.

\section{Numerical Simulations} \label{sec:sims}

In this section, we study how the evolution of $\omega$Cen and the GSE debris respond the stellar bar. For this, we need a reasonable estimate for the formation and evolution of the MW bar. The formation of the bar is expected to have begun $\gtrsim8\,$Gyr ago, based on observations of Mira variables in the nuclear stellar ring \citep[e.g.][]{sanders24}. The GSE itself is inferred to have merged with the MW earlier than this \citep[$8\text{--}11\,$Gyr ago, see][]{belokurov2018, haywood18b, helmi2018,orkney26}, meaning its debris must have been subjected to the entire lifetime of the Galactic bar from its onset.

Although the initial bar pattern speed and its long-term evolution are unknown, there are a number of constraints that we use to inform our models. By modelling the inner-region of the MW with made-to-measure models, \citet{portail17} derived a value of $\Omega_{b, 0}=39.0 \pm 3.5 \, \rm{km}\, s^{-1}\, kpc^{-1}$. On the other hand, \cite{sanders19} derived $\Omega_{b, 0}=41.0 \pm 3.0 \, \rm{km}\, s^{-1}\, kpc^{-1}$ using the continuity equation from the near side of the bar and $\Omega_{b, 0}=31.0 \pm 1.0 \, \rm{km}\, s^{-1}\, kpc^{-1}$ when also considering data from the far side. This is corroborated by \citet{bovy2019} and more recently more recently by \citet{khalil25} who reproduced all the fine details of the kinematics of the MW disc as measured from \textit{Gaia} DR3 in a dynamical model combining two spiral modes and a bar with a pattern speed of $37.0 \,\rm{km \, s^{-1} \,kpc^{-1}}$. Another study by \citet{horta25} reports much lower pattern speeds of $\Omega_{b, 0}=24 \pm 3 \, \rm{km}\, s^{-1}\, kpc^{-1}$. This value was already considered in \cite{dillamore2026}, where it is shown to naturally drag $\omega$Cen to its present-day location in $(E, L_{\rm z})$ space via resonance interactions, and so we do not consider it further here.

Galactic bars experience dynamical friction against the surrounding dark matter halo \citep{tremaine84,weinberg85, hamilton23} leading to angular momentum and energy loss.
This typically results in the bar slowing down and growing \citep{athanassoula92}. In the MW, \citet{chiba21} estimate that the rate of bar slow-down is $\dot{\Omega}_{\rm b}= 4.5 \pm 1.4 \,\rm{km \,s^{-1}\,kpc^{-1}\, Gyr^{-1}}$.

\begin{figure*}
\centering
  \setlength\tabcolsep{2pt}%
    \includegraphics[keepaspectratio, trim={0.2cm 0.0cm 0.2cm 0.0cm}, width=\linewidth]{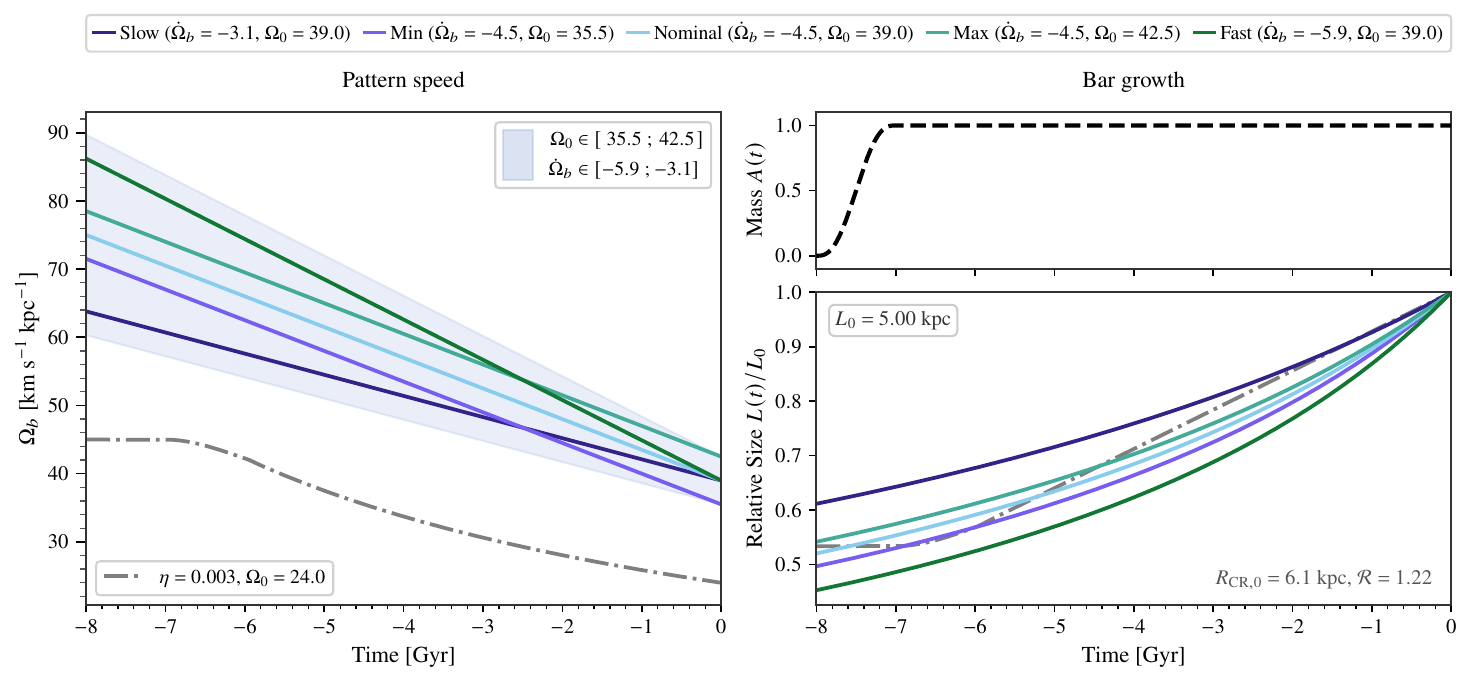}\\
\caption{Bar model parameters evolution. {\it Left:} Bar pattern speed evolution for the various models with varying degrees of slow-down rates $\dot{\Omega}_{b}$. The dotted-dashed grey curve is the model considered in \citet{dillamore2026}. {\it Top right:} Normalised mass growth of the bar. {\it Bottom right:} Bar size evolution.} 
\label{fig:bar_models}
\end{figure*}

\begin{figure*}
\centering
\setlength\tabcolsep{2pt}%
\includegraphics[keepaspectratio, trim={0.2cm 0.0cm 0.2cm 0.0cm}, width=\linewidth]{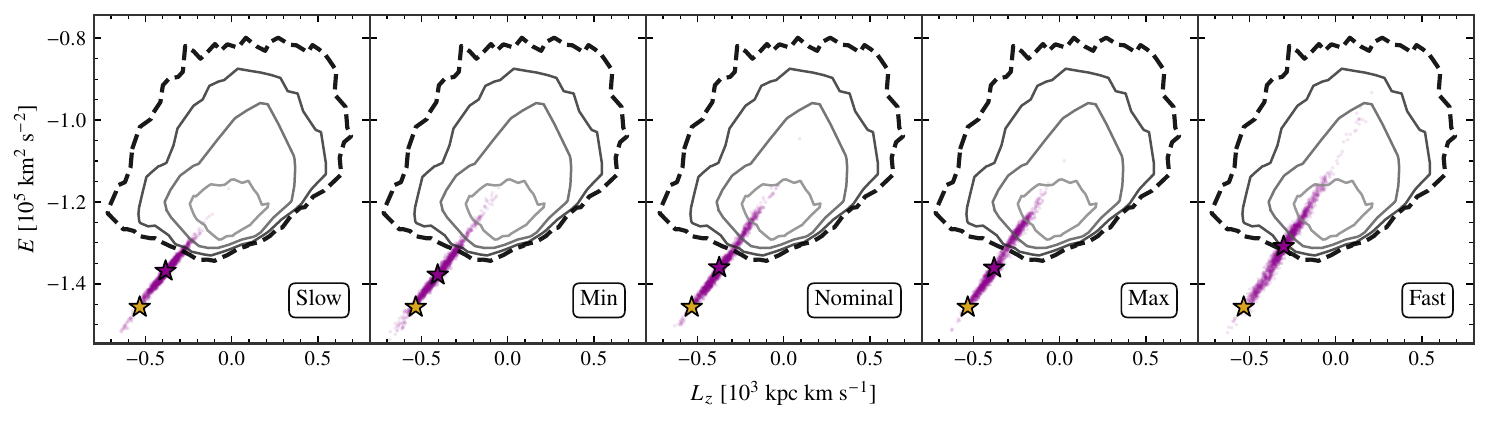}
\caption{ $\omega$Cen's migration in $(E, L_{\rm z})$ space for different bar pattern speed evolutions models. GSE debris contours of \citet{belokurov2023} are shown as the grey solid and black dashed lines. Present day location is shown the yellow star. Final distribution of backward integrated $\omega$Cen phase-space tracers is shown as purple dots. Past median $t=-8\, \rm{Gyr}$ location of $\omega$Cen is shown as a purple star. The appararent diffusion is a result of sampling $\omega$Cen's phase-space coordinate and its error.}
\label{fig:omegacen_migration}
\end{figure*}

\subsection{Setup}
\label{sec:setup}

We perform test-particle simulations of $\omega$Cen and a GSE-like population with {\sc agama} \citep{vasiliev19}, where we first use the MW potential of \citet{hunter24} that includes a decelerating stellar bar. We will later expand the analysis to the potential of \citet{khalil25}. These simulations are run over a period of 8\,Gyr, consistent with the expected age of the bar. Our goal is to examine whether resonances with the bar are capable of transporting stars from the GSE-like population to the low-energy and retrograde region of phase-space occupied by $\omega$Cen.

To this end, we consider 5 different models spanning the $1-\sigma$ minimum/maximum values of \cite{portail17}'s pattern speed (``Min'', ``Nominal'' and ``Max'')  with varying degrees of pattern deceleration rates (``Fast'' and ``Slow''). The bar is made to gradually increase in mass amplitude in the first billion years, while its size $L$ is made to evolve according to $L(t)=L_{0} \frac{\Omega_{b,0}}{\Omega_{b}(t)}$, where $L_{0}= \frac{R_{\rm{CR},0}}{\mathcal{R}}$ with $R_{\rm{CR},0}=6.1$\,kpc and $\mathcal{R}=1.22$ \citep{portail17}. This is illustrated in Figure~\ref{fig:bar_models}, where the left panel shows the bar's pattern speed as a function of time for the various models considered. To guide the eye, we also illustrate the model already covered in \citet{dillamore2026} as the dotted-dashed grey curve. On the right side, we show the evolution of the normalised mass of the bar (which is kept the same in all models) and the relative size evolutions for each model. Similar to \cite{dillamore2026}, we perform two types of simulations: a backwards integration of $\omega$Cen's phase-space position and a forward integration of particles drawns from a GSE-like distribution function (DF).   
\subsection{$\omega$Cen}
In our study we consider the phase-space position of $\omega$Cen as follows. Its proper motion is taken from \textit{Gaia} EDR3, distance from \cite{vasiliev21}, while its line-of-sight velocity and on-sky position is taken from \cite{baumgardt19} and \cite{harris10} respectively. We assume that the errors in distance and each velocity are Gaussian as reported in \cite{vasiliev21}. We set $\omega$Cen's phase-space position by taking $2\times10^{3}$ samples from the above distributions. We transform these into Galactocentric coordinates using the following convention $\mathbf{x}_{\odot}=(-8.178, 0, 0.028)$\,kpc and $\mathbf{v}_{\odot}=(11.1, 232.24, 7.25)\,\rm{km}\,\rm{s}^{-1}$ for the position and velocity of the Sun \citep[][]{schoenrich10, gravity19, bennet19, bovy2019}.

We then integrate these particles backwards to a time $t_{f}=-8\, \rm{Gyr}$. This is shown in Figure~\ref{fig:omegacen_migration} which traces the trajectory of $\omega$Cen's phase-space point in the energy versus angular momentum ($E,L_{\rm z}$) plane from the present day to its median location at $t_{f}=-8\,\rm{Gyr}$ shown by the yellow and purple stars respectively. It is clear that the bar is able to drag $\omega$Cen from its present day location in ($E,L_{\rm z}$) space to higher energies and angular momenta closer to $L_{\rm z}=0\,\rm{km\,s^{-1}\, kpc}$ to various degrees, including in some cases all the way inside the GSE debris region as defined by \cite{belokurov2023} using \textit{Gaia} DR3 data (shown as the grey and black dashed contours\footnote{These were rescaled to the new adopted potential of \citet{hunter24}.}). This is most evident for the ``Fast'' decelerating bar model.

\begin{figure*}
\centering
  \setlength\tabcolsep{2pt}%
    \includegraphics[keepaspectratio, trim={0.2cm 0.0cm 0.2cm 0.0cm}, width=\linewidth]{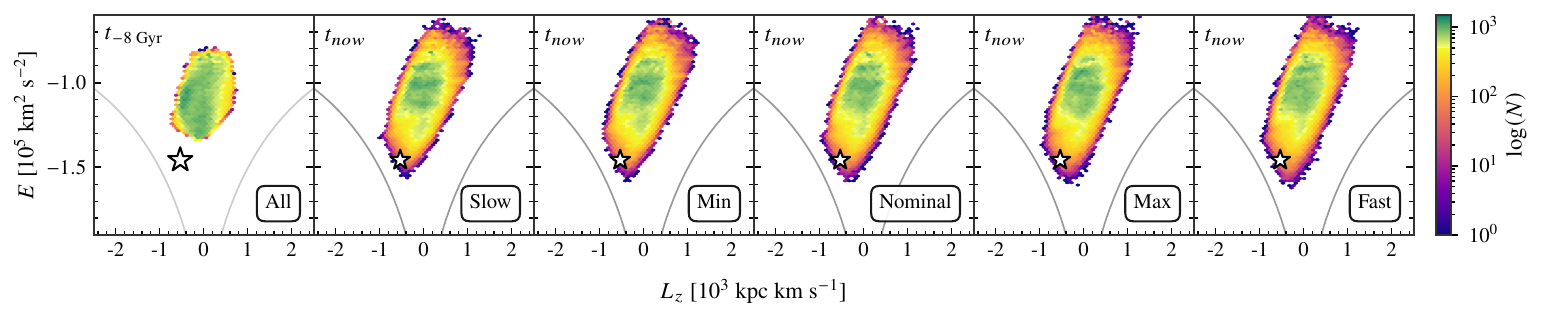}\\
\caption{Migration of GSE-like debris from $t=-8\,\rm{Gyr}$ to the present day for different bar pattern speed evolution models in ($E,L_{\rm z}$) space. $\omega$Cen's location is shown as the white star. In each of the models considered the GSE-like debris ends up reaching $\omega$Cen's present day location in ($E,L_{\rm z}$) space.
} 
\label{fig:GSE_migration}
\end{figure*}

We can also do the complementary exercise where we integrate forward a GSE-like DF sampled to fit within the contours of the GSE \citep{belokurov2023}, similar to what was done in \cite{dillamore2026} subject to the time-evolving barred potential from $t_{f}=-8\,\rm{Gyr}$ to the present day. The resulting $(E,L_{\rm z})$ distributions of this debris are shown in Figure~\ref{fig:GSE_migration}. We see that for each model considered, the evolved GSE debris naturally reaches the current phase-space location of $\omega$Cen in $(E,L_{\rm z})$ space. 

\subsection{Identifying the resonances at work}

\begin{figure*}
\centering
  \setlength\tabcolsep{2pt}%
    \includegraphics[keepaspectratio, trim={0.2cm 0.0cm 0.2cm 0.0cm}, width=\linewidth]{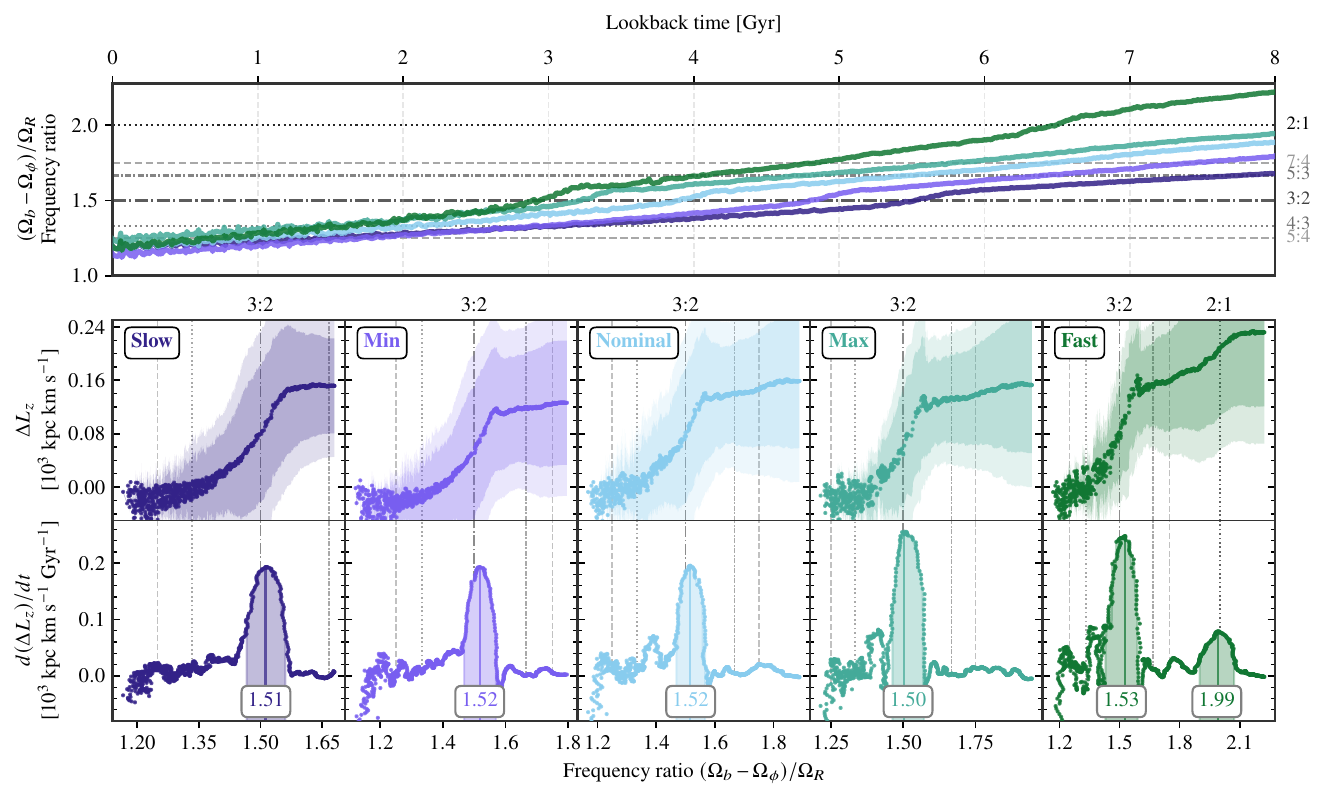}\\
\caption{ {\it Top panel:} $\omega$Cen's median trajectory through frequency space as a function of time for all 5 different models considered (coloured lines). The horizontal lines mark the different resonances sweeping through $\omega$Cen as a function of time. {\it Bottom panels}: Change and rate of change of angular ($\Delta L_{z}$ and $\dot{\Delta} L_{z}$ as a function of time for all 5 models. The most significant change of angular momentum occurs when the $3:2$ retrograde resonance crosses $\omega$Cen. In the ``Fast'' model there is also a non-negligeable contribution of the 2:1 retrograde resonance.  } 
\label{fig:resonance_interactions}
\end{figure*}

Our experiments show that the bar is capable of dragging radially anisotropic GSE stellar populations all the way to $\omega$Cen's current phase-space location at lower energies and retrograde orbits. However, the time evolving potentials considered here are complex, which result in different resonances sweeping along as a function of radius across the Galaxy. Thus, it is useful to identify which bar resonances are responsible in promoting $\omega$Cen's migration from the GSE to its present phase-space location. This is shown in Figure~\ref{fig:resonance_interactions}. The top panel follows $\omega$Cen's median trajectory evolution in frequency ratio $(\Omega_{b}-\Omega_{\phi})/\Omega_{R}$, for which different resonances sweep through it as a function of time. In the span of $\Delta t=8\,\rm{Gyr}$, different resonances cross paths with $\omega$Cen, however, only a select few provide efficient energy and angular momentum transfer. 

Indeed, closer inspection to the change of angular momentum $\Delta L_{\rm z}$ and its time derivative $\Delta \dot{L}_{\rm z}$ in the bottom panels of Figure~\ref{fig:resonance_interactions}, shows that the largest changes are mainly associated with the retrograde $3:2$ resonance, with the retrograde $2:1$ playing a minor role in some instances (e.g. ``Fast'' model).  This occurs at times when the bar pattern speed is equal to $\Omega_{b} \simeq60\,\rm{km\,s^{-1}\,kpc^{-1}}$. Depending on the model, the time spent at this resonance varies and is of the order $\Delta t_{3:2}/\approx0.5-1\, \rm{Gyr}$, small compared to the lifetime of the bar $t_{\rm bar}\sim8-10\, \rm {Gyr}$ \citep{sanders24, haywood24}.

\begin{figure}
\centering
  \setlength\tabcolsep{2pt}%
    \includegraphics[keepaspectratio, trim={0.2cm 0.0cm 0.2cm 0.0cm}, width=\linewidth]{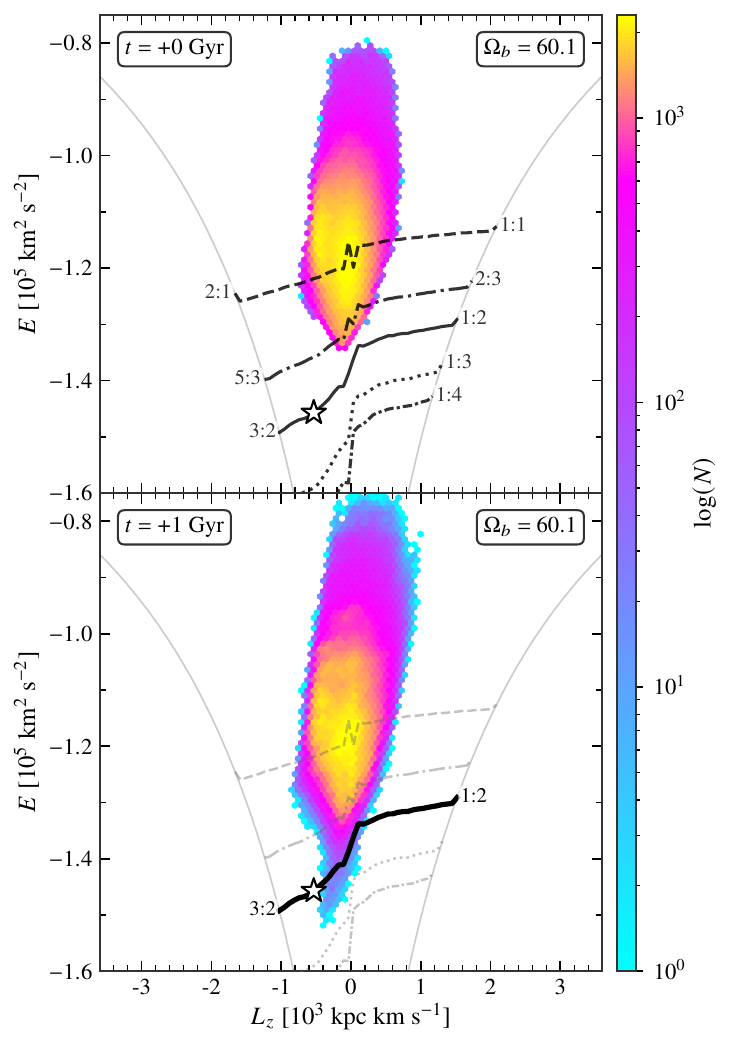}\\
\caption{ GSE-like stellar debris distribution in $(E,L_{\rm z})$ space in the \citet{hunter24} potential assuming a $\Omega_{b}=60\,\rm{km\,s^{-1}\, kpc^{-1}}$ pattern speed. {\it Top}: Initial condition of an anisotropic stellar halo within the GSE contours of \citet{belokurov2023}. Overlaid on top are the resonance lines. {\it Bottom}: Resulting distribution of GSE stars after applying the fixed bar patter speed potential of \citet{hunter24} for $\Delta t= 1 \,\rm{Gyr}$. Note the appearance of a trail of GSE stars following the $3:2$ retrograde resonance line reaching $\omega$Cen's position in ($E, L{\rm z}$).} 
\label{fig:GSE_resonances}
\end{figure}

To illustrate this better, we run a model keeping the bar pattern speed at $\Omega_{b}=60\,\rm{km\,s^{-1}\,kpc^{-1}}$. Figure~\ref{fig:GSE_resonances} shows the GSE-like particle distribution in ($E,L_{\rm z}$) space before and after forward time integration. We can see that for this particular pattern speed, the retrograde $3:2$ resonance goes through $\omega$Cen's current location and reaches the bottom of the GSE-like debris defined by \citet{belokurov2023}. After 1\,Gyr of evolution in this fixed pattern speed barred potential, the debris has already spread along the $3:2$ resonance ridge (shown as the solid black line) reaching energies and angular momenta consistent with $\omega$Cen's current location in ($E,L_{z}$) space.

\subsection{Dependency of the results on the assumed bar and Galaxy model}
\label{sec:yassin_model}

How general or sensitive are our results to the assumed bar model, Galactic model or modelling assumptions of the GSE-debris? This is the theme of this next section. Here we explore the sensitivity of our results to the assumed Galactic background model, to the bar model, and to the general modelling assumptions. For this, we adopt the bar potential first introduced in \citet{thomas23} and used in the non-axisymmetric Galactic model of \citet{khalil25} and its 3D generalisation in \citet{bougakov26}, also available in {\sc agama}\footnote{\url{https://github.com/GalacticDynamics-Oxford/Agama/blob/master/py/example_mw_potential_khalil25.py}}. The axisymmetric background of this model is described in \citet{khalil25}\footnote{We also adopt in this section a slightly different solar position and velocity to match \citet{khalil25}, namely $\boldsymbol{x}_{\odot} (R_\odot,\varphi_\odot,z_\odot)=(8.275, 0, 0.153) \, {\rm kpc}$ and velocity $\boldsymbol{v}_{\odot} = (V_{R_\odot},V_{\varphi_\odot},V_{z_\odot})=(-9.3, 251.5, 8.59) \, {\rm km}\, {\rm s}^{-1}$.}. Together with two spiral modes, the bar model reproduces the \textit{Gaia} DR3 RVS median radial velocity map across the Galactic plane, as well as much finer phase-space structure, recovering the disc moving groups both in the solar neighbourhood and across the disc. Only the bar is used here, and moreover it is made time-dependent with the same deceleration law as \citet{dillamore2026}, see section~\ref{sec:setup}, but with a different deceleration rate and different initial and final pattern speed $\Omega_b$. The initial $\Omega_b = 60\,\rm{km\,s^{-1}\,kpc^{-1}}$ starts decelerating at $t = -8.44$~Gyr down to its current $t=0$ value of $37\,\rm{km\,s^{-1}\,kpc^{-1}}$ \citep{khalil25}. The deceleration rate is  $\eta \equiv -\dot{\Omega}_b/\Omega_b^{2} = 0.0012$ \citep{chiba21}. The bar length scales with the corotation radius, $R_{\rm bar}(t) = R_{{\rm bar},0}\,R_{\rm CR}(t)/R_{{\rm CR},0}$, which for a flat rotation curve reduces to the scaling of section~\ref{sec:setup}. The orbital frequencies $\Omega_R$ and $\Omega_\phi$ are computed with {\sc agama} in the axisymmetric background.

\begin{figure}
\centering
  \setlength\tabcolsep{2pt}%
    \includegraphics[keepaspectratio, trim={0.2cm 0.0cm 0.2cm 0.0cm}, width=\linewidth]{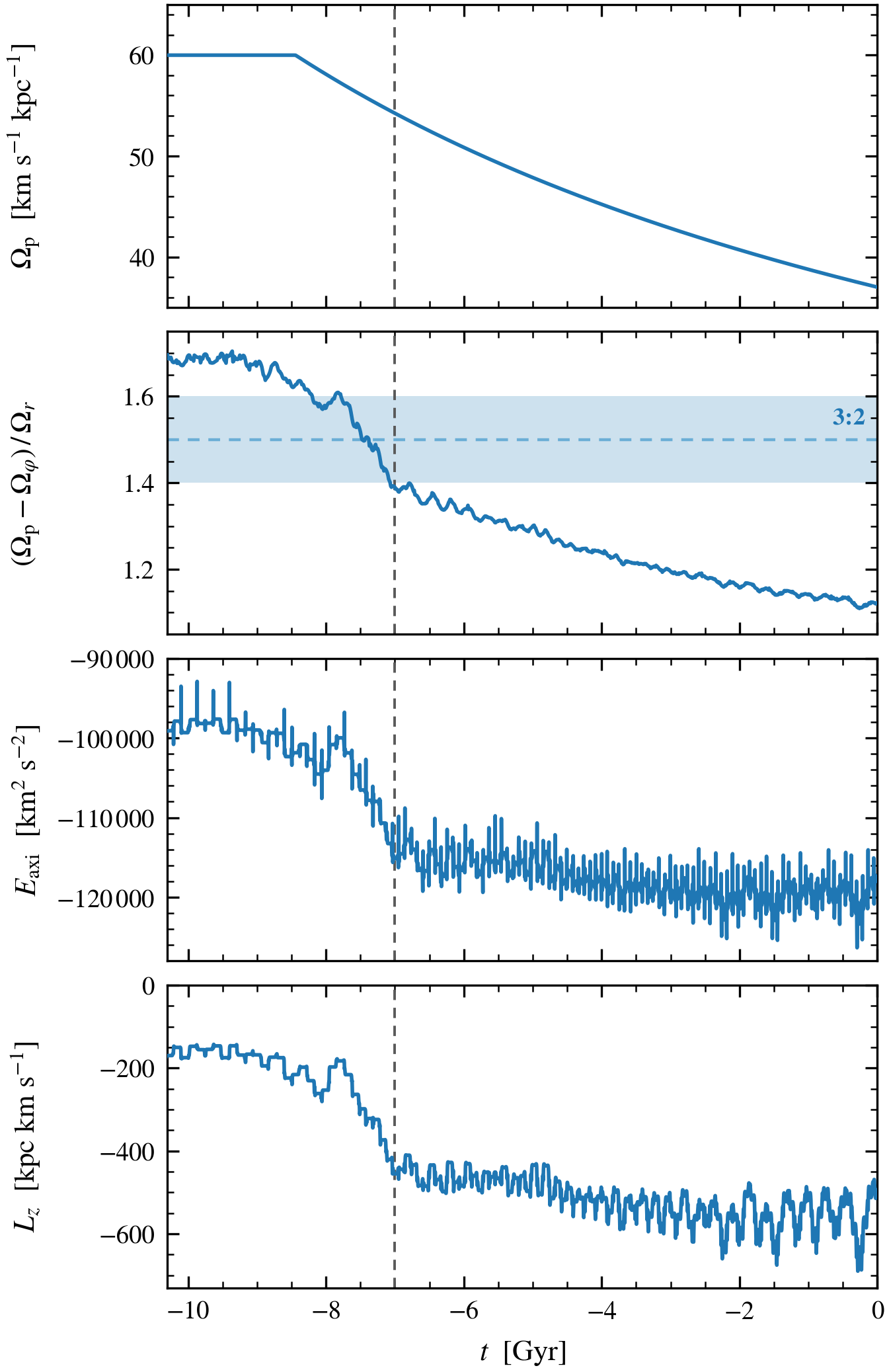}\\
\caption{
Adaptation of the \citet{khalil25} model, with bar pattern speed evolution and $\omega$Cen's evolution in frequency, energy and angular momentum in the top, top middle, bottom middle and bottom panels. Backwards integration starts at $t=0$ with a value $\Omega_{\rm p}\approx 37\,\rm{km \, s^{-1}\, kpc ^{-1}}$ and reaches $\Omega_{\rm p}= 60\,\rm{km \, s^{-1}\, kpc ^{-1}}$ by $t\approx - 8\,\rm{Gyr}$. $\omega$Cen crosses the $3:2$ retrograde resonance in between $t \approx -8$~Gyr and $t \approx -7$~Gyr, and sees rapid change in energy and angular momentum of order $\Delta E\approx20\,000\,\rm{km^{2}\,s^{-2}}$ and $\Delta L_{z}\approx 300\,\rm{km\, s ^{-1}\, kpc^{-1}} $ over $\Delta t\approx1\,\rm{Gyr}$ time-span. Note that the assumed Galactic potential and bar model is completely different from the \citet{hunter24} case and yet the conclusion is similar. The $3:2$ retrograde resonance thus offers a fairly generic energy/angular momentum transport mechanism. Regardless of the bar's pattern speed today (slow or fast), $\omega$Cen was likely affected by a bar resonance either retrograde $3:2$ or $1:1$.} 
\label{fig:yassin_model}
\end{figure}

In this model, the central value for the position and velocity of $\omega$Cen places it at $J_{\phi}=L_{z} = -494.2\,\mathrm{kpc\,km\,s^{-1}}$ and $E = -1.180 \times 10^{5}\,\mathrm{km^{2}\,s^{-2}}$, computed in the axisymmetric background potential. Since the energy zero-point also differs between the potentials, we compare the setups through the timing of resonance crossings, rather than through absolute positions in the $(E, L_{z})$ plane. The results are summarized in Figure~\ref{fig:yassin_model}, which shows the evolution of the bar's pattern speed, as well as $\omega$Cen's trajectory in frequency ratio space $(\Omega_b - \Omega_{\phi})/\Omega_R$, energy $E$, and angular momentum $L_{\rm z}$ as a function of time. As can be clearly seen, in this model, $\omega$Cen enters the $3:2$ retrograde resonance zone around $t\approx-7\,\rm{Gyr}$, which generates rapid change in energy and angular momentum of order $\Delta E\approx20\,000\,\rm{km^{2}\,s^{-2}}$ and $\Delta L_{z}\approx 300\,\rm{kpc\,km\,s^{-1}} $ on a timescale of $\Delta t\approx1\,\rm{Gyr}$. Despite the differences in assumed host potential and bar potential evolution, the conclusion therefore remains: the $3:2$ retrograde resonance offers an effective transport mechanism in energy and angular momentum space to dislodge $\omega$Cen from the GSE region and drag it to its current position in $(E,L_{\rm z})$ space. 

\subsection{A corridor between $\omega$Cen and GSE: bar model predictions}

\begin{figure*}
\centering
  \setlength\tabcolsep{2pt}%
    \includegraphics[keepaspectratio, trim={0.2cm 0.0cm 0.2cm 0.0cm}, width=\linewidth]{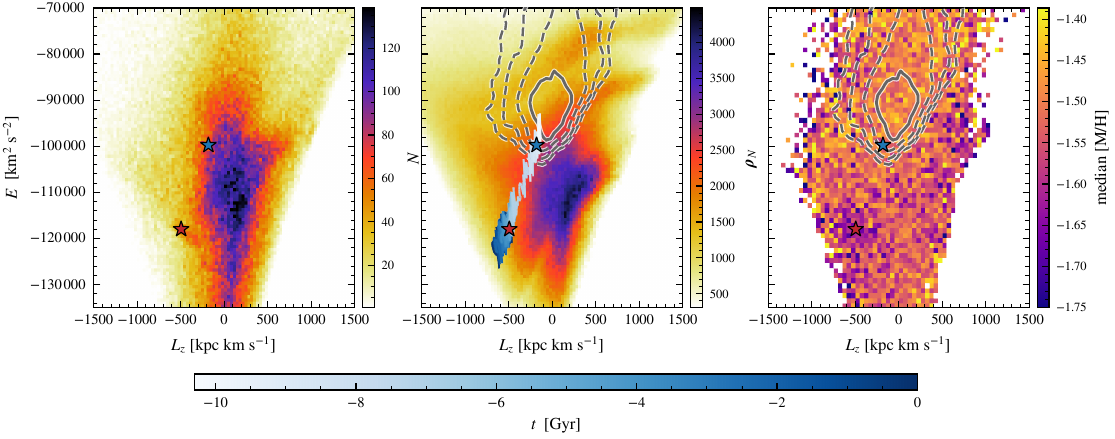}\\
\caption{$(E, L_{\rm z})$ distribution of the kinematically hot stars ($v_{\rm hot} > 220\,\rm{km\,s^{-1}}$). \textit{Left}: \textit{Gaia} DR3 $\times$ \texttt{StarHorse} star counts. \textit{Middle}: backward-integration model of the stellar halo in the decelerating barred potential of Figure~\ref{fig:yassin_model} (weighted DF, arbitrary normalisation). \textit{Right}: median [M/H] of \citet{andrae23}. The grey contours show the GSE debris of \citet{belokurov2023}, mapped onto the potential of \citet{khalil25}. The blue track is $\omega$Cen's orbit in the same model, coloured by time; the red star marks its present position and the blue star $t = -7.84\,\rm{Gyr}$.} 
\label{fig:yassin_ELz_model}
\end{figure*}

The action of the bar, particularly through different resonances, naturally promotes the migration of stars from high-energies to lower energies and in some cases changes in angular momentum in the retrograde sense. This should create a ridge \citep[e.g. see][]{dillamore2022, tomlinson26} but also a metal-rich corridor between the GSE and $\omega$Cen in $(E,L_{z})$ space. This would be comprised of proportionally metal-rich GSE and $\omega$Cen tidal debris. While current spectroscopic data is scarce, we can still present predictions of the DF based modelling and compare it to \textit{Gaia} DR3 data. Namely, ignoring GSE in the modelling, i.e. simply considering a slightly anisotropic stellar halo \citep{binney23}, we can explore how the present-day DF is distorted by the bar of subsection \ref{sec:yassin_model} in $(E-L_{\rm z})$ space compared to the data. This is shown in Figure~\ref{fig:yassin_ELz_model}. We also overlay the past trajectory of $\omega$Cen in the same model, and its positions at current time, $t=0$, and at $t=-7.84$~Gyr (subsection~\ref{sec:yassin_model}).

For the data, we use the \textit{Gaia} DR3 \citep{gaiadr3} stars with RVS line-of-sight velocities and \texttt{StarHorse} distances \citep{anders22}, transformed to Galactocentric coordinates with the solar parameters of subsection~\ref{sec:yassin_model} \footnote{We require $\texttt{ruwe} < 1.4$ \citep{lindegren21a, fabricius21}, an astrometric $\texttt{fidelity} > 0.5$ \citep{rybizki22}, $\varpi/\sigma_\varpi > 5$, and a relative distance precision $0.5\,(\texttt{dist84}-\texttt{dist16})/\texttt{dist50} < 0.2$, where \texttt{dist16}, \texttt{dist50} and \texttt{dist84} are the 16th, 50th and 84th percentiles of the \texttt{StarHorse} distance posterior.}. We then isolate the kinematically hot component with $v_{\rm hot} \equiv \sqrt{v_R^{2} + v_z^{2} + (v_\phi - v_{\rm c}(R))^{2}} > 220\,\rm{km\,s^{-1}}$, where velocities are measured relative to the local circular velocity $v_{\rm c}(R)$ of the axisymmetric part of the \citet{khalil25} potential, evaluated in the Galactic plane, leaving 274\,584 stars. The same velocity cut is applied to the model. The model prediction is obtained through backward integrating \citep{vauterin97, khalil25} within the time-dependent barred potential of subsection~\ref{sec:yassin_model}: the current value of the DF for each phase-space patch is the value of the initial DF for this patch integrated backward until before the bar existed. For the DF, we adopt the double power law in action space of \citet{binney23}, as distributed within {\sc agama}\footnote{Original implementation available in \url{https://github.com/GalacticDynamics-Oxford/Agama/blob/master/py/example_self_consistent_model_mw.py}, parameters described in \url{https://github.com/GalacticDynamics-Oxford/Agama/blob/master/data/SCM_MW.ini}}, with an anisotropy parameter $\beta=0.5$. We sample $2^{24}$ orbits, uniformly up to $4\,\rm{kpc}$ from the Sun, and the model map is then obtained as the integral of the DF over phase space in each bin of present-day $(E,L_{\rm z})$, applying the same $v_{\rm hot}$ cut as in the data. At low energies, the location and shape of the bar-induced ridges in the model are strikingly similar (at the qualitative level) to the data. Interestingly, some ridges predicted at higher energies do not seem to be present in the data at the same level, but this happens precisely in the GSE region, which is not included in our modelling.

Then, using the red-giant sample of \citet[][Sect.~4.2]{andrae23}, with [M/H] inferred from \textit{Gaia} XP spectra with XGBoost and selected by, among other criteria, $G < 16$, $\log g < 3.5$ and $T_{\rm eff} < 5200\,\rm{K}$, we draw the median metallicity map for $-1.9 < [\rm{M/H}] < -1.3$, and the same velocity cut \footnote{We also require $\varpi/\sigma_\varpi > 5$ and $G_{\rm BP} < 16.5$.} $v_{\rm hot} > 220\,\rm{km\,s^{-1}}$. This leaves $45\,319$ stars in the displayed $(E,L_{\rm z})$ region, where we show the median [M/H]  of the selected stars in bins with at least five stars. We do see a tentative hint for a metal-rich corridor carved out on the retrograde side of the bottom of the GSE debris contour in $(E,L_{\rm z})$ connecting to the $\omega$Cen.

\section{The stellar halo \& the $\omega$Cen-GSE connection}
\label{sec:stellar-halo}

Although current spectroscopic surveys of the MW do not provide enough sample statistics of high-resolution spectra (due to their pencil-beam nature) to confirm the existence of a bar-induced metallicity corridor connecting $\omega$Cen to the GSE, we are still able to probe other aspects of the stellar halo and NSCs. This is the aim of the following section for which we explore the structure of the stellar halo as well as the $\omega$Cen-GSE connection.

\subsection{Data}

For our analysis, we use the SDSS DR19 \citep{sdssdr19} catalog cross-matched with the \textit{Gaia} DR3 data \citep{gaiadr3}. We keep stars with distance errors $\sigma_{\rm dist}/d<0.2$. We prioritise distances derived by {\sc starhorse} and where not available we keep the inverse parallax distances from \textit{Gaia}. We also remove contaminants such as GCs, Sagittarius and the LMC which are masked using the membership catalogs of \citep{vasiliev21, cunningham24, jimenezarranz23} removing stars with $\mathcal{P}_i \geq 0.3$, where $i\in{\rm \{GC,\,Sgr,\,LMC\}}$.

In the next section, we will look at the chemical structure of the inner stellar halo.

\subsection{Properties of the MW's building blocks / inner halo}

In order to select inner halo stars we make additional cuts in Galactocentric radius and height, selecting stars with $R<30\,\rm{kpc}$ and $|z|<10\, \rm{kpc}$ and a metallicity cut of $\rm{[Fe/H]}\leq -1.0$. This ensures the removal of kicked-up disc stars \citep{dimatteo2019}. We further mask $\omega$Cen stars by excluding a 1\,degree ellipse around its physical location. \cite{hawkins2015} already recognised the power of using $\rm[Al/Fe]$ abundances as a means of differentiating accreted and in-situ stars and this was further refined in \cite{belokurov2022} to show how accreted/in-situ stars separate in $(E,L_{\rm z})$ space. In particular, they show how on average the in-situ population separates from the ex-situ population (i.e. the stellar halo, dominated by the GSE) with consistently higher [Al/Fe] abundances, from 0.5\,dex in the range $-1.3<[\rm{Fe/H}]<-0.9$ to 0.1--0.2\,dex at lower metallicities of $-1.7<[\rm{Fe/H}]<-1.3$. \cite{belokurov2022} define $[\rm{Al/Fe}]<-0.1$ as a cut for selecting accreted stars 
, albeit with some contamination contributions from high-$\alpha$ and low-$\alpha$ stars (see their figure 2). We impose a more conservative cut of $[\rm{Al/Fe}]<-0.2$ to minimise the contribution of the metal-poor disc $-1.5<[\rm{Fe/H}]<-1$ \citep{beers95}. We also additionally require \texttt{AL\_FE\_FLAG=0} and \texttt{FLAG\_BAD=0}.

\begin{figure}
\centering
  \setlength\tabcolsep{2pt}%
    \includegraphics[keepaspectratio, trim={0.2cm 0.0cm 0.2cm 0.0cm}, width=\linewidth]{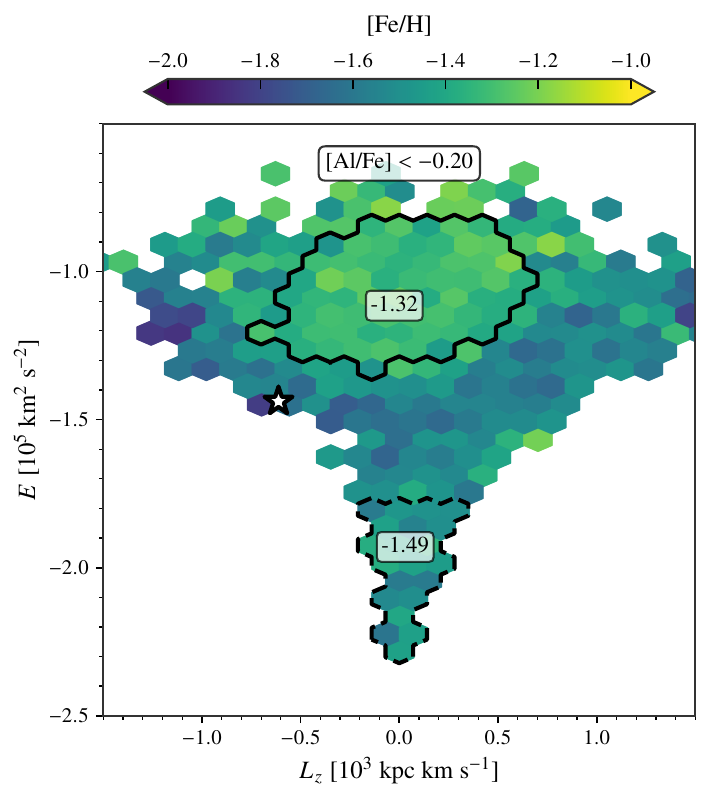}\\
\caption{Metallicity variations for ex-situ / building blocks components of the stellar halo for a selection cut in aluminium  ($[\rm{Al/Fe}]<-0.2$). GSE debris contours are shown as the black solid line. The stellar halo of the MW below the GSE debris is more metal-poor. This is characeterised by a ``canap\'e'' of $\rm[Fe/H]<-1.5$ material distributed just below the Sausage debris. The median metallicity of the MW's central building-blocks region is also more metal-poor than the GSE by 0.2\,dex. Due to its pencil beam nature, {\sc apogee} is not able provide enough statistics to confirm the presence of a metal rich corridor extending towards $\omega$Cen, this would need to await for future spectroscopic missions (e.g. 4MOST), however the metal-poor ``canap\'e'' provides extra support for a quiescent accretion history for the MW.}

\label{fig:metal_variations}
\end{figure}

The resulting ($E,L_{\rm z}$) map, colour-coded as a function of metallicity, is shown in Figure~\ref{fig:metal_variations} in the potential of \citet{hunter24}. The contribution of the GSE is still clearly evident. A new feature is also evident at lower energies: the presence of a metal-poor  ``canap\'e'' surrounding the GSE. While this feature was reported in \cite{laporte2026} using \textit{Gaia} XP spectra \citep{li2024}, here we are able to confirm it using high-resolution spectra. As we move to lower energies towards the innermost region of the MW's stellar halo, it becomes again more metal-rich, yet still more metal-poor than the GSE by $\sim0.2$\,dex. This tells us that whichever proto-Galactic `building blocks' that came before the GSE were less chemically enriched than the GSE itself. This possibly hints that the building blocks of the MW may have been less massive than the GSE.

\subsection{Chemistry of nuclear stars clusters and parent galaxies -- Sagittarius and implications for GSE and $\omega$Cen}


Chemically associating $\omega$Cen has been a source of contention in the past \citep[e.g.][]{anguiano2025, mason2026} due to the differing chemical enrichment trends seen in the GSE and $\omega$Cen, particularly pertaining to the existence of its secondary population. \cite{souza2026} argue that the sequence between \textit{Sequoia}, GSE, \textit{Thamnos} and $\omega$Cen show chemical continuity, however a closer inspection on the chemical trends shows that this is not the case. Indeed \cite{mason2026} use the same SDSS data to show that $\omega$Cen is not chemically compatible with any substructures that they considered (regardless of whether they are fictitious or real). Indeed, some stars in the first generation of $\omega$Cen have systematically higher [Al/Fe] than the bulk of GSE stars above a metallicity of [Fe/H]=-1.5. This gets even worse when considering $\alpha$-elements. Magnesium abundances in $\omega$Cen are shooting at an all time high value of $[\rm{Mg/Fe}]\gtrsim0.5$ -- there is simply no known field substructure in the Galaxy with such high abundances of $[\rm{Mg/Fe}]$ (GSE, \textit{Sequoia}, \textit{Thamnos} or any other substructures either real or fictitious). Thus, it is no surprise that there is no general chemical compatibility between $\omega$Cen and any other substructure in the MW. However, the GSE came with its own population of GCs (for which some could have undergone tidal stripping inside the GSE progenitor) and its own NSC. As such, there should still be a minute (but non-negligible) contribution of GC-origin stars in the GSE-``field'' originating from its associated NSC. We come back to this and discuss it further in section \ref{subsec: GC-origin stars}.

As noble and arduous of an endeavour chemically tagging $\omega$Cen to substructures in the field may be, and certainly kept many groups busy, let us take a step back and ask if general chemical compatibility is the legitimate proof for associating a NSC with its host galaxy (whether surviving or shredded). In other words, should there even be any hint of chemical semblance between field stars in a galaxy and its associated NSC in the first place? A galaxy forms the bulk of its field stars through clustered star formation, which for the majority form in loosely bound clusters that are subsequently dispersed \citep[e.g.][]{lada03, kamdar19}. Denser objects like GCs or a NSC forms within higher density environments with much more complex conditions \citep[e.g.][]{kravtsov05,elmegreen10,shapiro10, Kruijssen2012, hoyer26}. Clusters that form at the bottom of a galactic potential well become prone to mergers with other GCs that inspiral due to dynamical friction \citep[e.g.][]{tremaine75}. These are clearly not identical environments with different surface densities.

Nature is kind to us in the MW, providing us with an almost intact galaxy hosting a NSC. The surviving core of the Sagittarius (Sgr) dwarf galaxy \citep{ibata94} hosts at its centre a NSC: M54. Thus one can examine whether the secondary population in M54 is also seen in the Sgr core field stars.

In Figure~\ref{fig:M54_comparison}, we compare the [\rm{Al/Fe}] abundances of M54 with Sgr's core as well as those of $\omega$Cen with the GSE as a function of metallicity. Here, we require selecting giant stars using the following cuts:

\begin{itemize}
    \item $\log(g)<3.6$
    \item $3500<T_{\rm eff}<5500$
    \item $\texttt{SNR}>70$
    \item For M54: we select members with a probability $\mathcal{P}_{\rm M54}> 0.9$.
    \item For Sgr: $\mathcal{P}_{\rm Sgr}> 0.5$.
    \item For $\omega$Cen: $\mathcal{P}_{\rm \omega Cen}> 0.5$
    \item For GSE: belonging to the \citet{belokurov2023} contours minus known GC stars ($\mathcal{P}_{\rm GCs}> 0.3$).
\end{itemize}
While the trends in all $\alpha$-elements are similar between M54 and Sgr \citep[see e.g.][]{pagnini2025} -- this may be a coincidence due to low number statistics of APOGEE stars in M54 -- they certainly differ in their aluminium. Indeed, \cite{agaray2026} measured Mg and Al abundances for 45 stars in M54 spanning a metallicity range of $-1.5<[\rm{Fe/H}]<-1.0$, finding a strong anti-correlation in [Al/Fe]--[Mg/Fe] which is absent in Sgr's main body. On the other hand, looking at the distribution of Sgr's core stars, it is obvious that the ``galactic field'' is consistent with having no secondary population stars. Similarly, the presence of a single highly-enriched [Al/Fe] star does not signal field secondary population enrichment inside the core of Sgr, but rather the presence of stars originating from M54 liberated into the field. 

The differing trends in [Al/Fe] (and [Mg/Fe]) for NSCs and host galaxies is empirical evidence that secondary populations originate at a cluster-scale and not galaxy-scale level. This is substantiated by the absence of a secondary population in Sgr's own core, while it is clearly present in M54. Thus, the Sgr galaxy is living proof that secondary populations (so-called intermediate, P2 populations in $\omega$Cen) in GCs are cluster-scale physics dependent.

Similarly $\omega$Cen's chemistry does not require it to be chemically compatible with any general ``field'' chemical trends (not even in its first generation) in substructure, as confirmed by \citet{anguiano2025, mason2026}. Requiring this to be the case is ultimately a logical fallacy. 
Indeed, even when considering first generation stars in $\omega$Cen, the high-metal component of these stars can reach $[\rm{Mg/Fe}]\ge0.5$, values so high that they are not seen anywhere else in the stellar halo except for within other GCs \citep{schiavon24}. Finally, the apparent match between M54 and Sgr core field stars in $[\rm{Mg/Fe}]\text{--}[Fe/H]$ in Figure~\ref{fig:M54_comparison} may simply reflect the low number statistics of M54's secondary population stars. We demonstrate this by subsampling a random set of 17 stars in $\omega$Cen, which naturally picks up the more densely populated areas of the $[\rm{Mg/Fe}]\text{--}[Fe/H]$ plane and gives a false semblance of a chemical match in the general trend. Another interesting feature of Figure~\ref{fig:M54_comparison} worth noting, is the presence of high-$[\rm{Al/Fe}]$ ($[\rm{Al/Fe}]>-0.1$) GSE ``field'' stars branching out of the main trend. Some of these occupy regions similar to $\omega$Cen's secondary populations (intermediate and P2). These are likely GC-originating stars and potentially coming from $\omega$Cen stars. We explore these in \ref{subsec:GSE_debris_GC_stars}.

Recent work by \citet{agaray2026} suggest the chemistry of M54 is incompatible with galactic chemical evolution models or GC formation, instead preferring a formation via the merging of 2 GCs. This is similar to speculation about  $\omega$Cen. \citet{neumayer2020} argue that low-mass galaxies form NSCs primarily through GC inspiral, whereas high-mass galaxies do it via in-situ formation. Interestingly, the transition occurs at stellar masses of $\sim10^{9}\,\rm{M_{\odot}}$, which is in the ballpark that the GSE is predicted to have been. So, it is not implausible that the GSE formed a NSC through a messy combination of GC inspiral and in-situ formation, as with M54 (e.g. see references in \citealt{agaray2026}).

\subsection{GC originating stars in the GSE debris - a link to its ``missing'' nuclear star cluster}
\label{subsec:GSE_debris_GC_stars}

\begin{figure*}
\centering
  \setlength\tabcolsep{2pt}%
    \includegraphics[keepaspectratio, trim={0.2cm 0.0cm 0.2cm 0.0cm}, width=\linewidth]{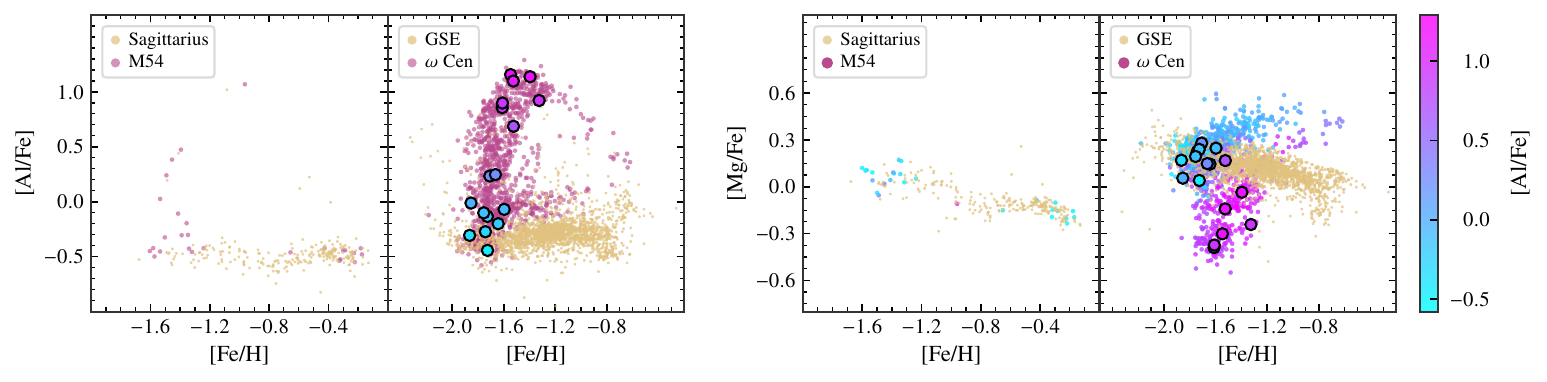}\\
\caption{Aluminium and magnesium abundance patterns as function of metallicity in host galaxy (golden points) and NSCs (red points). We additionally also highlight 17 randomly sampled $\omega$Cen stars. \textit{Left panels}: $[\rm{Al/Fe}]$ abundance patterns for the Sgr dwarf galaxy core and its central nuclear stars M54 as well as the GSE debris and $\omega$Cen juxtaposed together. Unlike M54 which shows a clear secondary population, Sgr shows no signs of a secondary population present in its core. From an empirical perspective this signals that secondary populations are not a galaxy-wide scale chemical enrichment phenomenon but rather an internal cluster-scale physics problem. Although the GSE and $\omega$Cen have similar levels of chemical enrichment in $[\rm{Fe/H}]$, the GSE has very little {\it general} overlap with $\omega$Cen. Note however, that the GSE displays a handful of stars with similar levels of $[\rm{Al/Fe}]$. These are GC-origin stars as we discuss in section \ref{subsec: GC-origin stars}, some of them belonging to $\omega$Cen. \textit{Right panels}: $[\rm{Mg/Fe}]$ abundance patterns for the Sgr dwarf galaxy core and its central nuclear stars M54 as well as the GSE debris and $\omega$Cen colour-coded by $[\rm{Al/Fe}]$.Again, $\omega$Cen's chemical pattern is visibly not compatible with the general field trend of GSE (but can seem to if poorly sampled, e.g. like M54 w.r.t to Sgr). We note again that the GSE exhibits a few stars with extremely low $[\rm Mg/Fe]$ abundances (similar to those in $\omega$Cen), signaling a likely GC origin.}

\label{fig:M54_comparison}
\end{figure*}

In the earlier section, we noted that many of the GSE debris stars have chemical patterns comparable to those exhibited by GCs (see Figure~\ref{fig:M54_comparison}). A large fraction of those same stars are not associated with any known bound GC, but do have metallicities compatible with the range covered by $\omega$Cen, leading us to consider the possibility that they originated within $\omega$Cen itself. To probe this possibility further, we leverage the fact that both GCs and NSCs harbour large populations of N-rich stars \citep[e.g. see][and references therein]{bastian18}, and the well-known trend that GCs exhibit a correlation in [Al/Fe] and [N/Fe] or [N/O] \citep[e.g. see][]{meszaros21,belokurov23b}. This makes N-rich stars an effective way to chemically tag field stars that originated in GCs \citep[e.g.][]{martell10, schiavon17, horta21,belokurov23b}. 

\begin{figure}
\centering
  \setlength\tabcolsep{2pt}%
    \includegraphics[keepaspectratio, trim={0.2cm 0.0cm 0.2cm 0.0cm}, width=\linewidth]{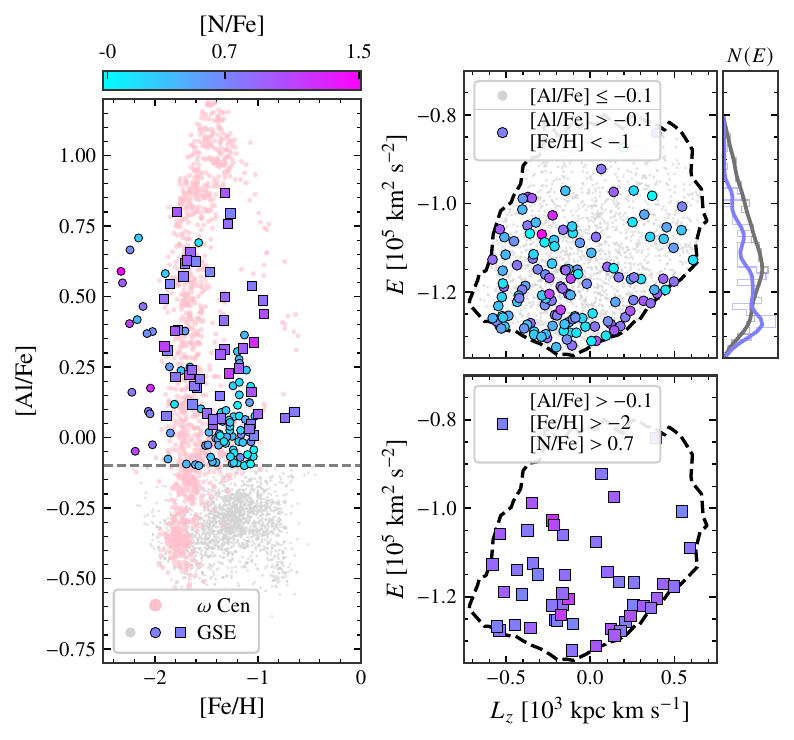}\\
\caption{Exploration of GSE debris high $[\rm{Al/Fe}]$ stars as defined by the \citet{belokurov2023} $E-L_{\rm z}$ contours. {\it Left panel:} [Al/Fe] vs [Fe/H] plane for GSE debris stars (grey points) and $\omega$Cen stars (red points). We highlight GSE debris stars with $[\rm{Al/Fe}]>-0.1$ as filled circles colour-coded by their $[\rm{N/Fe}]$ abundance. GSE debris stars with $[\rm N/Fe]>0.7$ with $[\rm{Fe/H}]>-2.0$ that are potentially $\omega$Cen compatible are represented as squares. {\it Top right panel:} Distribution of GSE debris stars with $[\rm Al/Fe]\leq-0.1$ and $[\rm{Al/Fe}]<-1.0$ shown as grey circles and $[\rm{N/Fe}]$ colour-coded circles in ($E,L_{\rm z}$) space. The GC-susceptible (potentially $\omega$Cen) origin GSE stars are more concentrated towards the lower energy end of the GSE debris than the ``field'' GSE stars. This is compatible with the idea that $\omega$Cen could have gotten dragged by the bar to lower energies. {\it Bottom right panel:} Distribution of N-rich ([N/Fe]>0.7) $\omega$Cen compatible ($-2.0<[\rm{Fe/H}]<-1.0$) GC-originating stars within the GSE contours. Again we notice that these stars cluster more towards the low-energy end of the GSE debris.}
\label{fig:GSE_debris_exploration}
\end{figure}

In Figure~\ref{fig:GSE_debris_exploration}, we present an exploration of GSE debris ``field'' stars contained with the \citet{belokurov2023} contour in ($E,L_{\rm z}$) space. On the right, we show the distribution of these stars as grey points in the [Al/Fe]--[Fe/H] plane. We also overlay $\omega$Cen's stars as red dots to guide the reader's eye. GSE stars with $[\rm{Al/Fe}]>-0.1$ (i.e. with abundances consistent with secondary populations in $\omega$Cen) likely to be of GC-origin are shown as circles and colour-coded by their nitrogen [N/Fe] abundance. Additionally, we highlight GC-origin ($[\rm{N/Fe}]>0.7$) GSE debris stars in the metallicity range $[\rm{Fe/H}]>-2.0$ consistent with the spread in metallicity of $\omega$Cen.

In the top right panel of Figure~\ref{fig:GSE_debris_exploration}, we compare the distribution of GSE field stars confined to the bulk of the GSE \citep[i.e. within the contours of][]{belokurov2023} with that of [Al/Fe]-rich stars (more GC-like) in ($E,L_{z}$) space. We also show on the side a kernel density estimation of the energy distribution for the ``field'' stars versus the ``GC'' susceptible stars. Interestingly, we see that the two distributions peak at different energies, with the GC stars ($\omega$Cen
compatible) peaking at lower energies than the bulk of the GSE field stars. In the bottom panel of Figure~\ref{fig:GSE_debris_exploration} we plot the distribution of N-rich stars ($[\rm{N/Fe}]>0.7$) and note that they also cluster towards the lower energy end of the GSE debris peanut. We take these as evidence that the GSE had a NSC residing at the bottom of the GSE's potential well, but also that it may have gotten dragged/dislodged, due to the difference in energy distribution between the field GSE stars and the GC-originating GSE stars. This motivates the design of an $\omega$Cen-compatible mask to chemically tag it to its associated galactic host debris in the field. This is explored in the next section.

\begin{figure*}
\centering
  \setlength\tabcolsep{2pt}%
    \includegraphics[keepaspectratio, trim={0.2cm 0.0cm 0.2cm 0.0cm}, width=\linewidth]{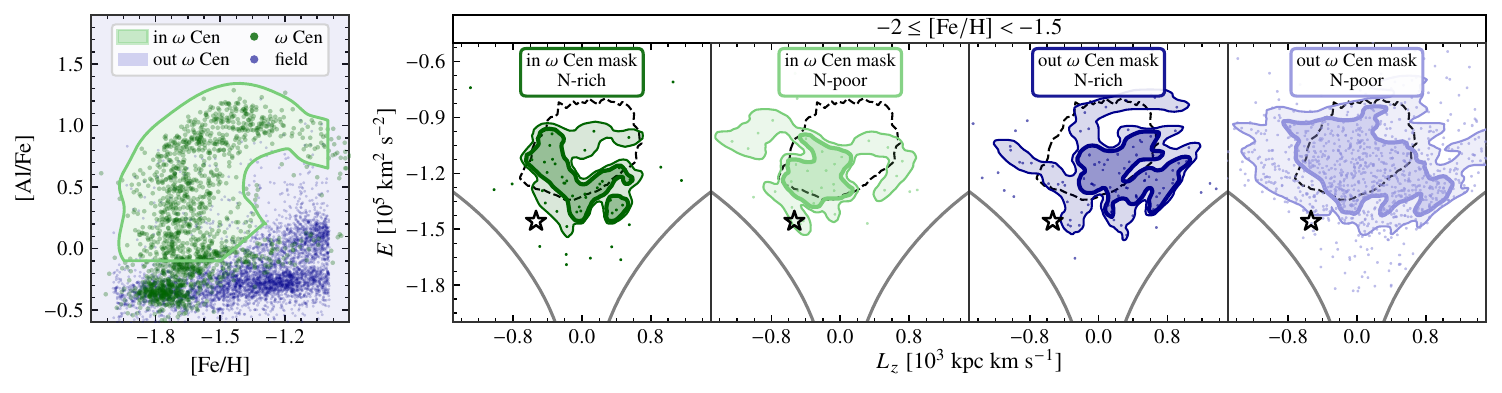}\\

\caption{ Our $\omega$Cen mask definition and corresponding distributions of N-rich / N-poor stars in ($E,L_{z}$) space originating from metal-poor GCs in the ranges $-2.0<[\rm{Fe/H}]<-1.5$. {\it Left Panel:} [Al/Fe] vs [Fe/H] plane showing $\omega$Cen stars and Galactic field stars (halo and in-situ stars in the range $-2.0<[\rm{Fe/H}]<-1.0$) as green and blue points respectively. The $\omega$Cen mask (green contour) is designed such as to minimise contamination from field halo and disc stars. Using this mask, we explore the distribution of stars that are compatible with $\omega$Cen in the range $-2.0<[\rm{Fe/H}]<-1.5$. This is a more restrictive metallicity cut to avoid as much disc contamination. We use stars outside of this mask as our ``control'' sample. {\it Right panels:} $(E,L_{\rm z})$ distributions of N-rich stars within the $\omega$Cen mask (dark green), N-poor stars within the $\omega$Cen mask (light green), N-rich outside the mask (dark blue) and N-poor stars outside the mask (light blue). The dashed contour denotes the location where the majority of the GSE debris is clustered from \citet{belokurov2023}. The $\omega$Cen compatible N-rich stars predominantly cluster at the lower-energy end of the GSE debris with a slight retrograde slant extending out to $\omega$Cen. A small fraction of stars are also present in the inner MW, likely contaminants originating from ancient proto-MW GCs. Similarly, the N-poor $\omega$Cen compatible stars occupy primarily a region consistent with the lower energy end of GSE, on its retrograde side predominantly also extending to $\omega$Cen's present day location in ($E,L_{\rm z}$). The control sample N-rich stars however, are more widely spread across $(E,L_{\rm z})$ space (both at higher and lower energies), with comparatively many more N-rich stars picked up about a wider range of circularities (also into the disc) than the $\omega$Cen compatible N-rich sample. The control sample of N-poor stars picks up stars in both the inner MW and across all the stellar halo, with no preferred sense of clustering.} 
\label{fig:nrich1}
\end{figure*}

\begin{figure}
\centering
  \setlength\tabcolsep{2pt}%
    \includegraphics[keepaspectratio, trim={0.2cm 0.0cm 0.2cm 0.0cm}, width=\linewidth]{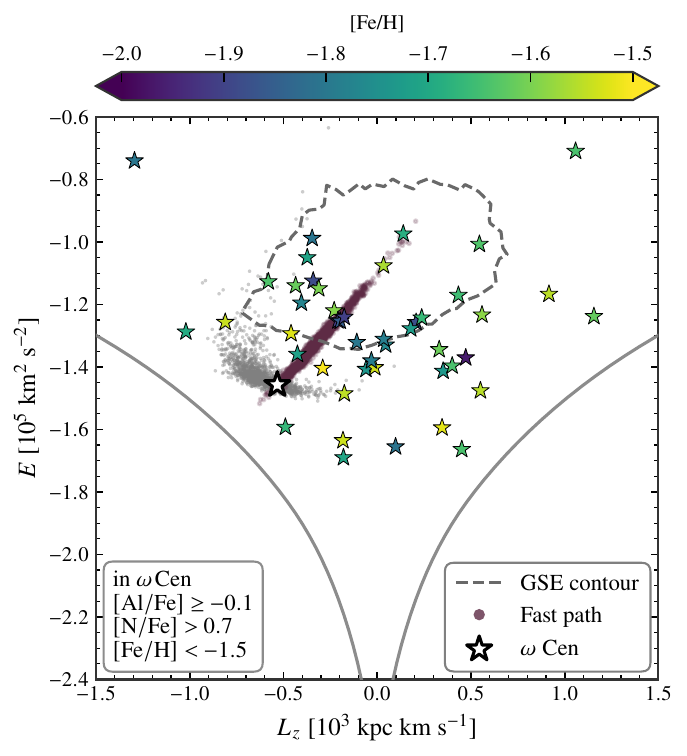}\\
\caption{Distribution of $\omega$Cen compatible N-rich stars colour-coded as a function of [Fe/H]. $\omega$Cen stars are shown as grey dots. We overlay the $\omega$Cen probability distribution for the ``Fast'' bar model which runs from $\omega$Cen's present day position towards the GSE debris. Note that the N-rich stars $\omega$Cen compatible cluster around the low-energy end of the GSE and in the direction of $\omega$Cen's present day location. We argue these are stripped stars from $\omega$Cen which got liberated and dragged along with it. These do not form a narrow trail for two reasons: stripping occurs at Lagrangian points, which have two different energies than the main body (this can be seen by the spread in $E,L_{z}$ of $\omega$Cen stars) and finally the action of the bar over time also creates additional diffusion in $(E,L_{z})$. This is the first independent high-resolution spectroscopically confirmed existence of the corridor joining GSE to the $\omega$Cen confirming the earlier results in \citet{laporte2026}.}
\label{fig:nrich2}
\end{figure}


\subsection{Probing the $\omega$Cen-Sausage GSE link with N-rich stars}
\label{subsec: GC-origin stars}


Here, we look for metal-poor N-rich stars in APOGEE DR19 \citep{sdssdr19} potentially associated with $\omega$Cen by selecting stars using the above mentioned cuts and imposing additional cuts and masks which we list below:
\begin{itemize}
    \item ${\rm log}(g)<3$,
    \item $T_{\rm eff}<5300\, K$,
    \item $\texttt{SNR} > 70$,
    \item $\texttt{N\_H\_FLAG}=0$, 
    \item $\texttt{AL\_H\_FLAG}=0$,
    \item a geometric mask onto Sgr with $|B_{\rm Sgr}|<10$ degrees about its orbital plane,
    \item a geometric mask of $15^{\circ}$ about the Magellanic Clouds,
    \item a mask on all GCs with a probability membership larger than $0.5$ from \cite{baumgardt19} and distance filterning\footnote{Around each cluster, a star is also removed if it lies both within an angular radius $\theta = \max\left(r_t/R_\odot,\; 1.1\,\theta_{99}\right)$ of the cluster centre and within the distance band $R_\odot \pm (r_t + 2\,\sigma_{R_\odot})$, where $r_t$ is the tidal radius, $R_\odot$ the heliocentric distance of the cluster and $\theta_{99}$ the 99th percentile of the angular separations of the catalogue members. The member-based term is capped at $3\,r_t/R_\odot$ and $\theta$ at $5^\circ$; the distance band is extended where needed to the 5th--95th percentiles of the member distances, and a star whose interval $d \pm 2\sigma_d$ overlaps the band is also removed.}.
    \item $\rm[N/Fe]>0.7$\footnote{We also double checked our results using $\rm[N/O]-\sigma_{\rm[N/O]}>0.55$ \citep{belokurov23b} and saw no appreciable difference with the results.},
    \item $[\rm{Al/Fe}]>-0.1$,
    \item $-2.0 <\rm{[Fe/H]}<-1.0$,
    \item an $\omega$Cen mask in $[\rm Al/Fe]\text{--}[Fe/H]$ shown in Figure~\ref{fig:nrich1}. 
\end{itemize}

The mask modulates the upper-envelope of $\omega$Cen's $[\rm Al/Fe]\text{--}[Fe/H]$ distribution. This selection is guided to pick GC-origin (not field) stars that have metallicities, $\rm[Al/Fe]$ and $\rm[N/Fe]$ abundances consistent with $\omega$Cen. The resulting distribution of these stars in $(E,L_{z})$ space is shown in Figure~\ref{fig:nrich1}. We also overlay the GSE debris contour of \cite{belokurov2023} to guide the eye. From this map, we can see that metal-poor N-rich stars show excess inside the GSE, particularly towards the lower energy end and region connecting to $\omega$Cen's present day location. Similarly the N-poor stars within the $\omega$Cen mask cluster predominantly close to the retrograde end of the GSE and also in the vicinity of $\omega$Cen. The control samples (N-rich or N-poor) on the other hand occupy much larger regions of phase-space from the inner-MW, to the disc and the stellar halo. The stars picked up by the $\omega$Cen masks clearly show a preferred sense of clustering according to how the GSE's NSC material should have been dragged from GSE towards $\omega$Cen by the bar. This is not coincidental because the chemical mask was applied on the full DR19 sample. 

Figure~\ref{fig:nrich2} shows the summary plot of $\omega$Cen compatible N-rich stars (shown as metallicity coloured stars) in $(E,L_{\rm z})$ space together with the distribution of $\omega$Cen stars (grey dots). We also overlay the probability distribution of $\omega$Cen from the ``Fast'' bar model backwards integration. This effectively gives an idea of the trajectory of $\omega$Cen potentially reaching the GSE centroid and connecting it to its present-day location. The fact the N-rich stars do not follow a thin narrow trailing path is two-fold. First, $\omega$Cen's stars which get stripped as it gets dragged by resonances become unbound through the Lagrangian points which have different energies (in fact the N-rich stars have about the same extent of spread as $\omega$Cen's stars shown as the grey dots). Second, the action of the bar also creates additional diffusion in ($E,L_{\rm z}$) as shown by \citet{dillamore2025}. 

This together with the results in Figure~\ref{fig:nrich1}, argues strongly for a link between $\omega$Cen and GSE, and in fact when folded with the action of the bar through retrograde resonances, establishes that the $\omega$Cen is the NSC of the Sausage GSE galaxy which got dislodged from it. Future spectroscopic surveys like 4MOST \citep{4MOSTs3}, including the 4DWARFS programme \citep{4DWARFs} will be able to cover much wider portions of the sky and potentially fully map $\omega$Cen's debris from its stream \citep[\textit{Fimbuthul}, see][]{ibata2019} to its more diffuse counter part which we unveiled \citep[see also][for an earlier chemical tagging study $\omega$Cen using barium abundances]{majewski12}.

\section{The rise and fall of the Sausage GSE galaxy and its nuclear star cluster: $\omega$Cen}
\label{sec:discussion}

In this section, we discuss the larger implications of our findings in the context of the formation of the MW.

$\omega$Cen is the most massive GC in the MW with properties akin to NSCs and long thought to be associated to an ancient destroyed galaxy. While many have speculated it to be associated to substructures in the MW \citep{meza2005, massari2019, myeong19, pagnini2025, souza2026}, no satisfactory dynamical or chemical explanation that would hold the line was ever provided, generally leading to more problems than solutions:

\begin{enumerate}
  
    \item If $\omega$Cen was associated to GSE, then why is it on a retrograde orbit and sitting at lower energies \citep[see][]{pfeffer2021} since dynamical friction is too weak and cannot circularise an orbit? \citet{souza2026} argued for chemical compatibility but could not explain this dynamical point.\\
    \item If $\omega$Cen is the central nucleus of a destroyed galaxy, why is it not chemically compatible with any known substructures \citep{anguiano2020, mason2026}? Indeed, since no component in the field of the MW has enrichment levels of $[\rm{Mg/Fe}]\sim0.5$, the question is a red herring.\\
    \item If $\omega$Cen is the NSC of another hypothetical galaxy ``\textit{Nephele}'' \citep{pagnini2025} or \textit{Sequoia} \citep{myeong19}, then where is the the NSC associated with GSE? Aside from M54, there simply is no other surviving lone unambiguous NSC in the MW\footnote{However, we note that \citet{pfeffer2021} argues that NGC 6273 could be a NSC potentially of Kraken/Heracles \citep[e.g.][]{Kruijssen2019, horta2021}}. This scenario raises another conundrum: why would $\omega$Cen survive long-term whilst the NSC from the GSE is entirely destroyed, despite them presumably having similar pre-infall masses and encountering similar tidal environments? Note that invoking too many mergers would also violate the constraint on the mass of the stellar halo \citep[e.g.][]{deason2019}.\\

\end{enumerate}

\citet{limberg2022} associated $\omega$Cen and other GCs to the GSE, but did not provide an explanation for its location in ($E,L_{z}$), yet argued that the GSE would still provide the best host given its speculated mass and implied mass for $\omega$Cen's IMBH. \cite{laporte2026} built on the intuition of \cite{limberg2022} and showed that the GCs associated to the GSE and $\omega$Cen's stellar populations followed the same age metallicity relation, both attaining similar limits of enrichment in [Fe/H]. They then identified bar resonances as a known mechanism capable of shifting both energy and angular momentum of the order of $\Delta E\approx0.4\times10^{5}\,\rm{km^{2}\, s^{-2}}$, $\Delta L_{z}\approx0.3-0.5\times10^{3}\, \rm{km \, s^{-1}\, kpc^{-1}}$ \citep[see also][]{tomlinson26} and uncovered a metal rich corridor connecting the GSE to $\omega$Cen which they interpreted as a smoking gun for this mechanism.

Our time evolving bar pattern speed models show that the $3:2$ retrograde resonance can naturally dislodge $\omega$Cen from the GSE debris' centroid within the span of $\lesssim 1\, \rm{Gyr}$. Moreover, this mechanism is general regardless of the chosen potential or bar model. The fact that this retrograde bar resonance mechanism works for all currently published present-day bar pattern speeds from the lowest \citep{horta25} to the highest \citep{portail17, sanders19,bovy2019, khalil25} values and is able to operate well within the lifetime of the bar is strong evidence that this process must haven taken place.

Our search for GC-originating N-rich stars that are chemically compatible with $\omega$Cen in APOGEE serendipitously uncovered a corridor of stars connecting $\omega$Cen all the way to the GSE with stars clustering strongly towards the low energy end of the GSE's debris all the way down to $\omega$Cen. 
It is remarkable that the prediction of $\omega$Cen's migration through $(E,L_{z})$ space lies roughly along the path of these N-rich stars, a testament of the predictive power of the model. We interpret these stars as the footprint trails of $\omega$Cen's resonant dragging through $(E,L_{z})$ space. While confirming the existence of a metal-rich corridor with field stars will need to await future spectroscopic surveys, our spectroscopically confirmed N-rich star sample provides the strongest tantalizing evidence yet that resonant dragging occurred somewhere in the past $10\text{--}8\,$Gyr.

\subsection{A consistent chemodynamical history of a significant merger from its outer edge to its inner-most nuclear star cluster}
Our results paint a novel and fully consistent history of events and dynamical mechanisms surrounding the evolutionary history of the last significant merger (the Sausage, GSE) from its outer region all the way to its NSC in tandem with that of our infant MW like never before. This also brings strong complementary constraints on its progenitor mass as well as timing of its merger using stars, GCs and $\omega$Cen's stellar populations as chronometers \citep{laporte2026}. These constraints shall only get tighter in the future with asteroresismic space missions such as PLATO and HAYDN.

It has now become evident beyond reasonable doubt that the MW suffered an ancient merger \cite[e,g,][]{chiba2000, brook2004, meza2005, deason13, belokurov2018, haywood18b, helmi2018} contributing to a large portion of the stellar halo in the form of a highly radially anisotropic structure. Such a configuration hints that this galaxy was massive enough to experience dynamical friction and eventually radialise during its final coalescence with the MW \citep{amorisco2017, fattahi2019, vasiliev2022}. The GSE's debris sits at relatively high energies \citep{belokurov2023,orkney26}; it does not dominate the centre of the MW. This points that the event was likely not more substantial than a $1:4$ merger \citep[see][]{orkney26}. Mergers of this nature occupy a large portion of phase-space in ($E,L_{\rm z}$) space \citep[see][]{jean-baptiste17, naidu2021, koppelman20} with strong variations in metallicity \citep{amarante2022}. This implies that many of the reported substructures such as {\it Sequoia}, {\it Thamnos} and {\it Pontus} \citep[][]{malhan2022} over the years which show too little differences between each other to be associated with distinct accretion events \cite{amarante2022, mori2024} are just belonging to a single entity. The association between GSE and \textit{Sequoia} is now almost evident both from a modelling side \cite{amarante2022} and data \citep{souza2026} as the outer parts of the GSE. From our analysis and implication of our models, \textit{Thamnos} is likely a misclassification of $\omega$Cen stars and GSE material, effectively sitting in the bar-carved corridor \citep{laporte2026}.

Spectroscopic field star subgiants from LAMOST \citep{xiang2022} place the onset of the last significant merger at $\sim11\,\rm{Gyrs}$ ago \cite{orkney26, laporte2026}. This is further corroborated by the existence of a starburst GC population {\it Tain\'a} with kinematic, chemical and age properties akin to centrally induced star-forming gas during mergers spanning anywhere between $1:10$ to $1:4$ \citep{orkney26}. The strong clustering of these GCs in relative age is consistent with a starburst event lasting a few 100\,Myrs \citep{laporte2026}. We can now also use $\omega$Cen as a chronometer for the shutdown of star formation in the GSE. Interestingly, the maximum age for its most metal-rich population is $11\,\rm{Gyr}$ \citep{clontz24}. When the GSE merged, the vast majority of its debris radialised to give the anisotropic stellar halo we see today \citep{belokurov2018} with $\omega$Cen at high energies close to $L_{z}\approx0\,\rm{km\,s^{-1}\,kpc^{-1}}$ \citep{belokurov2023}. As the bar formed shortly after $t\approx10-8 \,\rm{Gyrs}$ it trapped and dragged $\omega$Cen along a resonance line to lower energies and and retrograde part of $(E,L_{\rm z})$ space leaving on its way a trail of stripped N-rich stars from $\omega$Cen. This process should have equally affected the debris of the GSE giving rise to a metal rich corridor of GSE spilled material with it \citep{laporte2026}.

\subsection{Nuclear star cluster formation, Black hole formation, Dark Matter distribution, timing the Milky Way}

The association of $\omega$Cen to the galaxy responsible for the last significant merger, the GSE, now opens up new possibilities and avenues to study other aspects of galaxy formation through empirically motivated initial conditions, some of which we list below:

\begin{itemize}
    \item The formation of intermediate mass black holes (IMBHs) and their growth through gas flows.

    \item Chemical enrichment in NSCs within their dwarf galactic environments

    \item Dark matter distribution in the MW

    \item Constraining the various major life events of the Galaxy
\end{itemize}

Understanding the velocity DF of the dark matter distribution associated with the Sausage GSE and the final stages of stripping of the GSE will require folding in the effect of the growth and evolution of the bar into the modelling \citep[e.g.][]{dillamore2022, tomlinson26}. It also opens up the possibility of studying the ejection of dark matter in nuclear stars clusters with empirically motivated initial conditions. 

It also sets a lookback time upper limit for dynamical and chemical enrichment processes that took place in $\omega$Cen to give its peculiar chemical composition, through $t_{\rm \omega Cen_{\rm form}} < t_{\rm peri}=11 \, \rm{Gyr}$. The heated proto-disc stars (so-called ``Splash'') \citep{dimatteo2019, belokurov2020} gives information as to when the merger ended and thus gives an anchor to infer the merger duration $\Delta t_{\rm merger}$, currently situated at $\sim1-2\, \rm{Gyr}$ \citep{orkney26}. 

Finally, future HAYDN mission will be able to further constrain and infer ages of $\omega$Cen stars through asteroseismology and provide an extra anchor to timing the onset of the last significant merger and associated starburst $t_{\rm peri}\approx t_{\rm SB}$ but also in turn potentially constraining the period of star formation $\Delta t_{\rm SF}$ in the proto-MW and further refining the ages of the starburst induced {\it Tain\'a} GCs.

\section{conclusions} \label{sec:conclusion}

In this contribution we explored the effect of stellar bar resonances on $\omega$Cen as a potential catalyst for solving a long-standing problem on its origin by identifying its progenitor host galaxy. Our conclusions are as follows:

\begin{enumerate} 

    \item We show that the stellar bar can dislodge $\omega$Cen and drag it down to lower energies and retrograde orbit. This does not require a retrograde $1:1$ resonance for a slow bar \citep{horta25, dillamore2026}. The retrograde $3:2$ can also naturally explain this and arises for common bar pattern speeds \citep{portail17, sanders19, khalil25}. Hence, regardless of the bar pattern speed this mechanism is viable.\\

    \item Based on Sgr and M54's chemistry, we argue that searching for chemical compatibility between NSC and host galaxy (in the form of debris or intact galaxy) is a logical fallacy and ill-posed problem. No single debris in the stellar halo or any major MW component ever reach enrichment levels in [Mg/Fe] as high as 0.5 (except in GCs), and therefore cannot be compatible with the entirety of $\omega$Cen's populations.\\
    
    \item M54 empirically shows that the existence of different chemical populations in NSCs/GCs is a cluster scale specific physical problem, not a galaxy scale one.\\

    \item We spectroscopically confirm the existence of the $(E,L_{z})$ corridor connecting $\omega$Cen to the GSE by tracing N-rich $\omega$Cen-candidate stars back to the GSE centroid. The clustering of these stars is consistent with model expectations of resonant bar dragging. This confirms the early evidence presented in \textit{Gaia} XP spectra \cite{laporte2026} who interpreted this as a smoking gun for a bar resonance at the origin of $\omega$Cen's migration from the GSE. \\
    
    \item A model with a pattern speed consistent with $\Omega_{b}=39\pm3.5\,\rm{km\,s^{-1}\,kpc^{-1}}$ \citep{portail17, khalil25} naturally accounts for a dynamical connection between $\omega$Cen and the GSE, fits the trail of N-rich stars connecting the GSE to $\omega$Cen.\\ 
    
    \item Judging from \citet{dillamore2026}'s results, the $1:1$ retrograde prediction is also consistent with our uncovered N-rich distribution of stars clustered about the line joining $\omega$Cen and the GSE centroid in ($E,L_{\rm z}$).\\  

    \item All these elements unambiguously demonstrate that $\omega$Cen is the remnant nucleus of the Sausage (GSE) galaxy.\\

    \item \textit{Thamnos} is likely to be an example of a fictitious substructure, being an amalgam of GSE material that was dragged onto a more retrograde orbit by bar resonances, and stripped $\omega$Cen stars.\\
   
\end{enumerate}
Our study brings a solution to a $\sim$40 year old puzzle on the origin of $\omega$Cen for which we made both the prediction and found observational evidence that affirms it with public data from {\it Gaia} and SDSS DR19. This opens a new avenue towards understanding the formation of NSC in their galactic hosts, the distribution of dark matter in the Galaxy and growth of IMBHs.

\section{Acknowledgements}
 CL acknowledges useful discussions with Danny Horta on GCs and their chemistry. CL \& MO acknowledge funding from the European Research Council (ERC) under the European Union’s Horizon 2020 research and innovation programme (grant agreement No. 852839). CL also acknowledges funding from the Agence Nationale de la Recherche (ANR project ANR-24-CPJ1-0160-01).  YRK acknowledges funding from the European Research Council (ERC) under the European Union’s Horizon 2020 research and innovation programme (grant agreement No.~101117455). BF acknowledges funding from the Agence Nationale de la Recherche (ANR-25-CE31-4684). YRK acknowledges ISCRA for awarding this project access to the LEONARDO supercomputer, owned by the EuroHPC Joint Undertaking, hosted by CINECA (Italy). CL thanks Thayenne for inspiring the ideas that led to this work.
  
\section*{Data Availability}

Simulation data, and any post-processed data that is unique to this work, can be made available to the enquirer upon request.



\bibliographystyle{mnras}
\bibliography{references} 



\appendix

\bsp	
\label{lastpage}
1\end{document}